\documentclass[superscriptaddress,onecolumn,pre,nofootinbib]{revtex4}

\usepackage{amsmath}
\usepackage{graphicx,color}

\newcommand{\be}{\begin{equation}}
\newcommand{\ee}{\end{equation}}
\newcommand{\bea}{\begin{eqnarray}}
\newcommand{\eea}{\end{eqnarray}}

\begin{document}
\sloppy


\title{Kinetic theory of spatially homogeneous systems with long-range interactions: \\
II. Historic and basic equations}

\author{Pierre-Henri Chavanis}
\affiliation{Laboratoire de Physique Th\'eorique (IRSAMC), CNRS and UPS, Universit\'e de Toulouse, F-31062 Toulouse, France}

\begin{abstract}

We provide a short historic of the early development of kinetic theory in plasma physics and synthesize the basic kinetic equations describing the evolution of systems with long-range interactions derived in Paper I. We describe the evolution of the system as a whole and the relaxation of a test particle in a bath of field particles at equilibrium or out-of-equilibrium. We write these equations for an arbitrary long-range potential of interaction in a space of dimension $d$. We discuss the scaling of the relaxation time with the number of particles for non-singular potentials. For always spatially homogeneous distributions, the relaxation time of the system as a whole scales
like $N$ in $d>1$ and  like $N^2$ (presumably) or like $e^N$
(possibly) in $d=1$. For always spatially inhomogeneous distributions, the relaxation
time of the system as a whole scales like $N$ in any dimension of space.
For 1D systems undergoing a dynamical phase transition from a homogeneous to an inhomogeneous phase, we expect
a relaxation time of the form $N^{\delta}$ with $1<\delta<2$ intermediate between the two previous cases.
The relaxation time of a test particle in a bath always scales
like $N$.  We also  discuss the kinetic theory of systems with long-range
interactions submitted to an
external stochastic potential. This paper gathers basic equations that are applied to
specific systems in Paper III.

\end{abstract}

\maketitle

\section{Introduction}
\label{sec_introduction}

The kinetic theory of systems with long-range interactions is currently a topic of active research \cite{houches,assise,oxford,cdr,bgm,proceedingdenmark}. Initially, kinetic theories were developed for 3D plasmas and 3D stellar systems. However, they are now considered at a more general level for different types of potentials of interaction in different dimensions of space. It is therefore desirable to develop a general formalism valid for arbitrary potentials with long-range interactions and, from that formalism, consider different applications. This is the purpose of this series of papers. In Paper I \cite{paper1}, using the Klimontovich approach, we have derived the basic kinetic equations governing the evolution of spatially homogeneous systems with long-range interactions (these equations may be derived  equivalently from the BBGKY hierarchy). In Paper III \cite{paper3}, we apply them to specific systems. In particular, we consider systems with power-law potentials (including plasmas and stellar systems in one, two, and three dimensions) and
the Hamiltonian Mean Field (HMF) model with attractive or repulsive interactions. In order to facilitate the applications, we summarize in this paper the basic equations and provide additional results that were not given previously. Although these equations have been established for a long time in plasma physics for the 3D Coulombian interaction, no synthesis of these equations and of their generalization to other potentials exists yet. We think therefore that the synthesis given in this paper may be useful to many readers working in the field. We also provide a short historic of the early development of kinetic theory in plasma physics because it is in that context that the basic kinetic equations (Landau, Vlasov, Lenard-Balescu) were first derived. This historic is certainly incomplete
but it contains references to important contributions that do not seem to be well-known.
It may therefore help reconstructing in detail the body of
knowledge of this topic.

The first kinetic theory was developed by Boltzmann (1872) \cite{boltzmann} for
a dilute neutral gas in which the atoms or the molecules have short-range
interactions. He obtained a kinetic equation by using simple semi-intuitive
arguments. If we apply it to systems of particles interacting with power-law
potentials decreasing at large distances as $r^{-\gamma}$ we
find that collisions with large impact
parameters produce a divergence when $\gamma\le 1$ in $d=3$.
The critical exponent $\gamma_c=1$ precisely corresponds to the
Coulombian interaction. Therefore, the kinetic theory of neutral gases developed
by Boltzmann is not well-suited to handle inverse-square forces between ionized
particles in a plasma. In the classical theory of neutral gases \cite{chapman},
the assumption is made that only the relatively close encounters are important
and that the forces between particles at greater distances have no effect. This
is precisely the inverse situation that is met in a plasma. Distant encounters,
producing small deflections, are more important than close encounters. It is
therefore necessary to develop a specific kinetic theory for plasmas.

The first kinetic equation of plasma physics was derived by Landau (1936) \cite{landau} who considered the effect of binary collisions between electrons and ions in a spatially homogeneous Coulombian plasma. He started from the Boltzmann equation and made a weak deflection approximation. Indeed, for the Coulomb potential, slowly decreasing with the distance like $r^{-1}$, weak collisions are the most frequent ones (the probability of large scattering is small). Each encounter induces a small change in the velocity of a particle but the cumulated effect of these encounters leads to a macroscopic process of diffusion and friction that drives the system towards statistical equilibrium. It is therefore possible to expand the Boltzmann equation in terms of a small deflection parameter $\Delta\ll 1$ and make a linear trajectory approximation. This treatment yields a logarithmic divergence at small and large impact parameters but the equation can still
be used successfully if appropriate cut-offs are introduced. The divergence
at small scales arises because the Landau theory, based on a weak coupling
approximation, does not take proper
account of the processes occurring in close binary collisions where the
trajectories of the particles are no more rectilinear.  A
natural lower cut-off $\lambda_L$, which is called the Landau length, corresponds
to the impact parameter leading to a deflection at $90^{o}$. The divergence at large scales
arises because the Landau theory does not take collective effects into account.  Phenomenologically, the Debye length\footnote{The Debye length was first introduced in connection with screening processes in highly ionized electrolytes by Debye and H\"uckel (1923) \cite{dh1,dh2}.  Each particle of the plasma tends to attract to it particles of opposite charge and to repel particles of like charge, thereby creating a kind of ``cloud'' of opposite charge. This cloud screens the electric field produced by the particle on a distance of the order of the Debye length.} provides a natural  upper cut-off. Indeed, beyond the Debye length  $\lambda_D$ many particles interact and the electric field of a particle is screened by its neighbors. Therefore, the particles do not interact if their separation exceeds the Debye length. In this sense, the interaction is effectively short-ranged.

Since the Coulombian interaction between charges shares many analogies with the Newtonian interaction between masses, it is useful to discuss in parallel the kinetic theory of self-gravitating systems.
Chandrasekhar (1942,1943a,1943b) \cite{chandra,chandra1,chandra2}
developed a kinetic theory of stellar systems in order to determine the
timescale of collisional relaxation and the rate of escape of stars
from globular clusters.\footnote{Early estimates of the relaxation time of stellar systems were made by Schwarzschild (1924) \cite{schwarzschild}, Rosseland  (1928) \cite{rosseland}, Jeans (1929) \cite{jeansbook}, Smart (1938) \cite{smart}, and Spitzer (1940) \cite{spitzerevap}. On the other hand, the evaporation time was first estimated by Ambartsumian (1938) \cite{ambart} and Spitzer (1940) \cite{spitzerevap}.} To simplify the kinetic theory, he considered
an infinite and homogeneous system and studied the relaxation of a test particle
in a thermal bath of field particles.
He started from the Fokker-Planck equation and determined the diffusion coefficient and
the friction force (second and first moments of the velocity
increments) by considering the mean effect of a succession of two-body
encounters. This approach was based on Jeans (1929) \cite{jeansbook} demonstration that the cumulative effect of the weak deflections resulting from the relatively distant encounters is more important than the effect of occasional large deflections produced by relatively close encounters.  Since his approach can take large deflections into account, there is no divergence at small impact parameters and the
gravitational analogue of the Landau length appears naturally in the
treatment of Chandrasekhar. However, his approach leads to a
logarithmic divergence at large scales that is more difficult to remove in stellar dynamics than in plasma physics because of the absence of Debye shielding for the gravitational force. In a series of
papers, Chandrasekhar and von Neumann (1942) \cite{cvn} developed a completely
stochastic formalism of gravitational fluctuations and showed that the
fluctuations of the gravitational force are given by the Holtzmark
distribution (a particular L\'evy law) in which the nearest neighbor
plays a prevalent role. From these results, they argued  (see also Jeans \cite{jeansbook} and Spitzer \cite{spitzerevap}) that the
logarithmic divergence has to be cut-off at the interparticle distance $l\sim n^{-1/3}$
where $n$ is the numerical density of stars. However, since the interparticle
distance is smaller than the Debye length, the same arguments should
also apply in plasma physics, which is not the case. Therefore, the
conclusions of Chandrasekhar and von Neumann are usually taken with
circumspection. In particular, Cohen {\it et al.} (1950) \cite{cohen} argued that the logarithmic divergence should be cut-off at the Jeans length since the Jeans length is the gravitational analogue of the Debye length. Indeed, while in neutral plasmas the effective interaction distance is limited by the Debye length, in a self-gravitating system, the distance between interacting particles is only limited by the system's size.   Chandrasekhar (1949) \cite{nice} also
developed a Brownian theory of stellar systems and showed that, from a
qualitative point of view, the results of kinetic theory can be
understood very simply in that framework as a competition between diffusion and friction.\footnote{The stochastic evolution of a star is primarily due to many small deflections produced by relatively distant encounters. The problem of treating particles undergoing numerous weak deflections was originally encountered in relation to the Brownian motion of large molecules which are thermally agitated by the smaller field molecules. The stochastic character of the many small impulses which act on a suspended particle is usually described by the Fokker-Planck equation. Chandrasekhar  \cite{chandrastoch} (1943) noted the analogy between stellar dynamics and Brownian theory and employed a Fokker-Planck equation to describe the evolution of the velocity distribution function. Indeed, he argued that a given star undergoes many small-angle (large impact parameter) collisions in a time small compared with that in which its position or velocity changes appreciably. Since the Newtonian potential makes the cumulative effect of these small momentum transfers dominant, the stochastic methods of the Fokker-Planck equation should be more appropriate for a stellar system than the Boltzmann approach. However, the coefficients of diffusion and friction are finally calculated with the binary-collision picture. This is equivalent to expanding the Boltzmann equation for weak deflexions (as in the Landau theory) since the Fokker-Planck equation can precisely be obtained from the Boltzmann equation in that limit.} In particular, he showed
that a {\it dynamical friction} is necessary to maintain the
Maxwell-Boltzmann distribution of statistical equilibrium and that the
coefficients of friction and diffusion are related to each other by an
Einstein relation which is a manifestation of the fluctuation-dissipation theorem. This relation is confirmed by his more precise
kinetic theory based on two-body encounters. The Fokker-Planck
approach of Chandrasekhar was further developed  by Rosenbluth {\it et al.} (1957)
\cite{rosen} who proposed a simplified derivation of the coefficients of diffusion
and friction valid for an arbitrary distribution function. It is important to
emphasize, however, that Chandrasekhar  did not derive the kinetic
equation for the evolution of the system as a whole. Indeed, he
considered the Brownian motion of a test star in a fixed
distribution of field stars (bath) and derived the corresponding
Fokker-Planck equation.\footnote{Indeed, Chandrasekhar
\cite{chandra1,chandra2} models the evolution of
globular clusters by the Kramers equation which has a fixed
temperature (canonical description) while a more relevant kinetic
equation would be the Landau equation  which conserves the energy
(microcanonical description).} King (1960) \cite{kingL} noted that if we
were to describe the dynamical evolution of the cluster as a whole,
the distribution of the field particles should evolve in time in a
self-consistent manner so that the kinetic equation must be an
integrodifferential equation. The kinetic equation obtained by King turns out to be
equivalent to the Landau equation, although written in a different
form.\footnote{Apparently, the Landau equation was not known by astrophysicists at that time who rather followed the treatment of Chandrasekhar \cite{chandra,chandra1,chandra2} and Rosenbluth {\it et al.}
\cite{rosen}. To our knowledge, the first explicit reference to the Landau equation in the astrophysical literature appeared in the paper of Kandrup (1981) \cite{kandrup1}. There is also a strange comment related to the work of Landau in the paper of Cohen {\it et al.} (1950) \cite{cohen} that suggests that his work was not clearly
understood.} There is, however, an important
difference between stellar dynamics and plasma physics. Neutral
plasmas are spatially homogeneous due to electroneutrality and Debye shielding.  By
contrast, stellar systems are spatially inhomogeneous. The
above-mentioned kinetic theories developed for an infinite homogeneous
system can be applied to an inhomogeneous system only if we make a
{\it local approximation}. In that case, the collision term is
calculated as if the system were spatially homogeneous or as if the
collisions could be treated as local. Several astrophysicists  have then
 tried to incorporate effects of spatial inhomogeneity in the kinetic theory of stellar systems (see further discussion in \cite{aanew}).

Returning to plasma physics, the early works on kinetic theory were based on the model of binary collisions
treated either by the Boltzmann or, more adequately, by the Fokker-Planck equation. However, these theories
lead to difficulties because the collision term diverges at large impact parameters. This is due to the
fact that the Coulomb potential does not decrease rapidly enough at large distances, which means
physically that collective interactions, involving many particles simultaneously, are important in
determining the behavior of the system. In plasma physics, this divergence has been suppressed by
introducing rather arbitrarily a cutoff in the integration at the Debye length which roughly
measures the effective screening of the interaction by the polarization
of the medium.\footnote{Before Landau (1936) \cite{landau},
several authors like Gruner (1911) \cite{gruner}, Chapman (1922) \cite{chapmanplasma},
and Persico (1926) \cite{persico} attempted to develop a kinetic theory
of ionised gases to model the interior of giant stars. Gruner \cite{gruner} applied the kinetic theory
of gases to power-law potentials
$r^{-\gamma}$ and noted that a divergence occurs when $\gamma=1$. Chapman \cite{chapmanplasma} proposed
to truncate the diverging integral at the interparticle distance (the Debye shielding was not yet discovered).
 Persico \cite{persico}, following the works of Rosseland \cite{ross} and Eddington \cite{edd}, understood that the
divergence could be solved by taking collective effects (Debye shielding) into account. He
replaced the Coulombian potential of interaction by the Debye potential in the expression
of the diffusion coefficient coming from the theory of gases so that no divergence occurs at large scales.} However, this
{\it ad hoc} procedure is not satisfactory and a proper treatment of collective
effects must be developed. Indeed, the encounters are clearly not binary since
there are many particles inside the screening sphere.

A first treatment of collective effects in plasma physics was introduced by Vlasov (1938,1945) \cite{vlasov1,vlasov2}. He treated collective effects in a self-consistent manner but neglected ``collisions'' between charges. Therefore,  each charge is described as moving under the influence of an average field due to all the others. The evolution of the distribution function is then governed by the collisionless Boltzmann equation coupled to the Poisson equation. This  system of equations had been previously introduced by Jeans (1915) \cite{jeans} for stellar systems under the same approximations (see H\'enon (1982) \cite{henon} for interesting historical details). However, the Vlasov equation cannot be used for a description of transport processes. Because of the neglect of collisions, it contains no mechanism of relaxation toward thermal equilibrium.\footnote{There is, however, an irreversible phenomenon of Landau damping (1946) \cite{landaudamping} which is a process of uniformization in configuration space. In stellar dynamics, the Vlasov-Poisson system may also undergo a process of violent collisionless relaxation that has been discussed by Lynden-Bell (1967) \cite{lb} in terms of statistical mechanics. Violent relaxation is generally incomplete and sometimes leads to ``core-halo'' states (see \cite{levin} and Appendix B of \cite{stan} for a detailed discussion).}

Pines and Bohm (1952) \cite{pb} developed for the first time  a description of electron interactions taking collective effects and binary collisions into account. They showed that for phenomena involving distances greater than the Debye length, the system behaves collectively while for distances shorter than the Debye length, it may be treated as a collection of approximately free individual particles, whose interaction may be described in terms of two-body collisions. This justifies therefore the Landau procedure of cutting-off the integration at the Debye length. They studied screening properties of out-of-equilibrium plasmas. For particles at rest or moving sufficiently slowly, they showed that the field is screened as in the Debye-H\"uckel theory. However, if particles are moving with a speed greater than the mean thermal speed, the field propagates away from the source as a wake. As a result, an excess of particles of like-charge accumulates ahead of the moving particle and conversely an excess of particles of opposite-charge accumulates behind it. Under such circumstances, polarization occurs and the potential field becomes asymmetrical. They concluded that the interaction between electrons in a plasma can be described in terms of short-range collisions between ``effective free particles'' corresponding to the electrons plus their associated cloud.

Bogoliubov (1946) \cite{bogoliubov} was the first to improve the Landau
kinetic theory by taking collective effects into account. Starting from the
Liouville equation and using  what is now called the ``BBGKY hierarchy'' and the
``Bogoliubov ansatz'', he derived a kinetic equation in which the collision term
is expressed in terms of a two-body correlation function that is
the solution of an integral equation (this equation corresponds
to the first term of an expansion
in powers of the coupling parameter $g=1/(n\lambda_D^3)$ so it is valid in a ``weak coupling'' approximation). In
this equation, the Debye shielding is automatically taken into account so that the integral over the space
coordinates is convergent at large distances contrary to the treatment of Landau who had to introduce a cutoff
artificially at the Debye length in order to make a logarithmically divergent integral finite. In addition, while the Boltzmann and Fokker-Planck equations used by Landau and Chandrasekhar have a phenomenological character, the  method of Bogoliubov, that starts from the Liouville equation, is systematic and rigorous. However, the final equation for the evolution of the distribution function given by
Bogoliubov is not explicit.

Lenard (1960) \cite{lenard} solved the Bogoliubov integral equation exactly by using a Fourier transformation and obtained an explicit kinetic equation where the time derivative of the distribution function is expressed explicitly in terms of the distribution function itself. The same kinetic equation, describing the collective effects in a weak coupling approximation, was obtained independently by Balescu (1960a,1960b) \cite{balescu1,balescu2} from the general theory of irreversible processes in gases developed by Prigogine and Balescu (1957) \cite{priba}. In this formalism, one starts from the Liouville equation, perform a Fourier transformation and solve the obtained equations by an iteration procedure. The asymptotic evaluation of the orders of magnitude of the various contributions to the Fourier components is accomplished by means of a diagram technique. This leads to an integral  equation equivalent to the  Bogoliubov integral equation that  Balescu  solved by using essentially the same method as Lenard.
This is valid in a weak coupling approximation at the order $g=1/\Lambda$ where $\Lambda=n\lambda_D^3$ is the number of charges in the
Debye sphere. The Lenard-Balescu kinetic equation describes the interaction of ``quasi particles'', which are electrons or ions ``dressed'' by their polarization clouds (this picture in terms of ``effective free particles'' -electron plus its associated cloud- was previously suggested by Pines and Bohm). These clouds are not a permanent feature, as in equilibrium theory, but they have a nonequilibrium changing shape distorted by the motion of the particles. The Lenard-Balescu equation exhibits a new type of nonlinearity which is directly related to the collective nature of the interactions. Besides the product of two distribution functions $f(1)f(2)$ characteristic of any two-body collision term, it contains the distribution function in the denominator. In fact, the Lenard-Balescu equation differs from the original Landau equation by the appearance of the dielectric function $|\epsilon({\bf k},{\bf k}\cdot{\bf v})|$ in the denominator of the potential of interaction $\hat{u}(k)$. Physically, this means that the particles are ``dressed'' by their polarization cloud. The Landau equation is recovered when $|\epsilon({\bf k},{\bf k}\cdot{\bf v})|$ is replaced by unity, {\it i.e.} when collective effects are neglected. However, with this additional term, it is found that the logarithmic divergence that occurs at large scales in the Landau equation is removed and that the Debye length appears naturally. In the dominant approximation, the Lenard-Balescu equation reduces to the Landau equation with a large-scale cut-off at the Debye length (which is now rigorously justified).

The Lenard-Balescu equation describes the interaction of the electrons which are ``dressed'' by a polarization could.\footnote{This notion of ``dressed test particles'' is further developed by Rostoker (1964) \cite{rostoker}.} This effect is well-known in equilibrium as the Debye cloud. However, in the situation considered by Lenard and Balescu, the polarization cloud differs sensibly from the latter because the system is out-of-equilibrium and the polarization cloud is deformed by the motion of the particles. For small velocities the deformation is merely a flattening in the direction of the motion but for velocities larger than the thermal velocity the cloud degenerates into a wake of plasma oscillations. This precisely accounts for what Pines and Bohm  have shown in a more heuristic manner. For small velocities, or in the dominant approximation, the effect of screening in the Lenard-Balescu equation is equivalent to replacing the bare Coulomb potential by the Debye potential (in agreement with Pines and Bohm). Interestingly, Liboff (1959) \cite{liboff} had previously developed a kinetic theory of plasmas with a Debye potential in order to avoid divergences at large scales.\footnote{Liboff computes the diffusion and friction coefficients
from the two-body encounters theory taking strong collisions into account, so he has no divergence at small scales neither.} He showed that, in the dominant approximation, the Debye potential gives the same results as the ``Coulomb plus Debye cutoff''.

The Lenard-Balescu equation was also derived independently by Guernsey (1960) \cite{guernseyphd} so it is sometimes called the Lenard-Balescu-Guernsey equation.\footnote{The kinetic theory of Guernsey, and other kinetic theories, are presented in the book of Wu (1966) \cite{wu}.} The works of Lenard, Balescu, and Guernsey have removed the problematic divergence at large scales that appears in the Landau equation. However, their approach based on a weak coupling approximation yields a divergence at small scales because it does not take  strong collisions into account. Coulombian plasmas therefore seem to be described by two complementary kinetic theories, depending on the value of the impact parameter $\lambda$: (i) the {\it binary encounters (or impact) theory} based on the Boltzmann or Fokker-Planck equations which is appropriate to describe strong collisions when  $\lambda\sim\lambda_L$ (it does not present any divergence at small scales but yields a logarithmic divergence at large scales) and (ii) the {\it wave theory} based on the Lenard-Balescu equation that is appropriate to describe collective effects when  $\lambda\sim\lambda_D$ (it does not present any divergence at large scales but yields a logarithmic divergence at small scales). In the intermediate case  $\lambda_L\ll \lambda\ll\lambda_L$, these two kinetic theories reduce to the Landau theory (it presents divergences both at small and large scales). Baldwin (1962) \cite{baldwin}, Frieman and Book (1963) \cite{fb}, Weinstock (1964) \cite{weinstock}, and Guernsey (1964) \cite{guernsey}  proposed a kinetic equation that takes  strong collisions and collective effects into account and exhibits no divergence.  Their treatment essentially amounts to writing the collision term as a sum of two terms: a Lenard-Balescu term that takes collective effects into account and a Boltzmann term that takes strong collisions into account. This leads to a general equation combining the Boltzmann and the Lenard-Balescu equation. Another unification scheme has been proposed by Kihara and Aono (1963) \cite{ka} by introducing an artificial splitting of the integration over the impact parameter in the binary encounters theory and over the wavenumber in the wave theory. Finally, Gould and DeWitt \cite{gould} (1967) obtained a convergent kinetic equation by using a direct analogy with the equilibrium theory of the electron gas. They also carefully compared their results with those of previous works.

Prior to Lenard and Balescu, several authors like Temko (1957) \cite{temko}, Kadomtsev (1958) \cite{kadomtsev}, Tchen (1959) \cite{tchen}, Ichikawa (1960) \cite{ichikawa}, and Willis (1962) \cite{willis} had attempted to solve the Bogoliubov integral equation. However, they solved it in an approximate manner which amounts essentially to making the Debye-H\"uckel approximation, {\it i.e.} to using the equilibrium form of the correlation function. Although their treatment is approximate, it shows that the correlations that exist in equilibrium are sufficient to provide both short and long distance cutoffs. Furthermore, the treatment of Tchen goes beyond the Debye-H\"uckel approximation and captures the ellipsoidal anisotropy of the shielding cloud first evidenced by Pines and Bohm  for a moving particle.  These authors obtained explicit kinetic equations that have the form of the Landau equation in which the Coulombian potential of interaction  is replaced by the Debye-H\"uckel potential (or more general potentials). This justifies the procedure of Liboff  who introduced the Debye-H\"uckel potential phenomenologically. In the dominant approximation, the velocity dependence of the dielectric function arising in the Lenard-Balescu equation may be neglected, so these heuristic treatments are sufficient for 3D plasmas.

In parallel, several authors like Gasiorowicz {\it et al.} (1956) \cite{gasiorowicz}, Temko (1957) \cite{temko},
Tchen (1959) \cite{tchen}, Rostoker and Rosenbluth (1960) \cite{rr}, Thompson and Hubbard (1960) \cite{th}, and
Hubbard (1960a,1960b)  \cite{hubbard1,hubbard2} developed a kinetic theory of homogeneous plasmas for a test
particle in a thermal bath, taking collective effects into account. They showed that the large-scale divergence
that appears in the Landau diffusion coefficient and friction force is actually regularized at the Debye length. In particular,
Hubbard \cite{hubbard1,hubbard2} considered the relaxation of a test particle in a thermal bath (one particle out-of-equilibrium moving in a gas that is in equilibrium), or in a bath that is not necessarily in thermal equilibrium, and derived the diffusion and friction coefficients entering in the Fokker-Planck equation. He showed that the diffusion is due to the fluctuations of the electric field while the friction is due to a polarization process.\footnote{This interpretation is particularly clear in the work of Gasiorowicz {\it et al.} \cite{gasiorowicz}. As a particle moves through the distribution of particles, the latter becomes polarized because the test particle attracts the field particles, so that they tend to concentrate behind it, and thus the force from particles behind is larger than the force due to those in front, with the result that the test particle is slowed down.} The field produced by the test particle polarizes the plasma. The charge displacement in the plasma representing this polarization is highly correlated with the position of the test particle and gives rise to an electric field which acts on the test particle and produces a friction. His calculations take into account the effect of correlations between distant electrons or ions. Consequently, the Debye screening is automatically included in his kinetic theory and it is not necessary (as it had been in the earlier calculations) to introduce a cut-off procedure to suppress the effects of distant encounters between particles. His calculations also include a proper treatment of close binary encounters like in the standard Boltzmann theory described by
Chapman and Cowling \cite{chapman}. As a result, he  obtains Fokker-Planck coefficients which include both collective effects and the contribution of close binary encounters. In his  theory, no {\it ad hoc} cut-off procedures of any kind are needed. In the dominant approximation, he recovers the results of Landau but now without any divergence. The Landau and Debye lengths appear naturally in his treatment. It is interesting to note, for historical reasons, that the results obtained by  Hubbard  are very closely related to those obtained independently by Lenard and Balescu at the same time. Indeed, if we substitute the diffusion and friction coefficients found by Hubbard in the Fokker-Planck equation and perform minor transformations (a substitution that Hubbard did not explicitly make), one obtains the Lenard-Balescu equation!

In the following years, alternative derivations of the Lenard-Balescu equation have been given. Some derivations are based on the BBGKY hierarchy but use a procedure different from that of Lenard and Balescu. For example, Dupree \cite{dupree1} and Ichimaru \cite{ichimaru} solve  the second equation of the BBGKY hierarchy in terms of operators (propagators) which are related to the solution of the linearized Vlasov equation. On the other hand, Nicholson \cite{nicholson} solves the second equation of the BBGKY hierarchy by using Fourier-Laplace transforms (while Lenard and Balescu use only Fourier transforms). Other derivations  due to Dupree  \cite{dupree2}, Fried \cite{fried}, Klimontovich \cite{klimontovich}, and Lifshitz and  Pitaevskii \cite{pitaevskii} start from the  Klimontovich equation and use a quasilinear approximation. All these methods are equivalent but the calculations are presented in a different manner. It is usually acknowledged that the approach based on the Klimontovich  equation is
technically simpler (see Paper I).

Recently, it was realized by the community working on systems with long-range interactions that these kinetic equations have a wider scope than just plasma physics and stellar dynamics. Actually, they are valid for arbitrary potentials of long-range interactions, attractive or repulsive, in a space of dimension $d$. Although the form of the Lenard-Balescu equation is unchanged (we just need to replace the Fourier transform of the Coulombian potential by the Fourier transform of the more general potential $\hat{u}(k)$), the results (relaxation time, diffusion coefficient, polarization cloud...) sensitively depend on the form of the potential of interaction and
on the dimension of space. It is therefore useful to expose general kinetic equations valid for arbitrary potentials with long-range interactions, then consider different applications (Paper III).

\section{The thermodynamic limit and the weak coupling approximation}
\label{sec_scaling}

We consider a system of material particles in interaction in a $d$-dimensional space. Their dynamics is described by the Hamilton equations
\begin{equation}
m\frac{d{\bf r}_i}{dt}=\frac{\partial H}{\partial {\bf v}_i},\qquad m\frac{d{\bf v}_i}{dt}=-\frac{\partial H}{\partial {\bf r}_i}
\label{wsm1}
\end{equation}
with the Hamiltonian
\begin{equation}
H=\sum_{i=1}^N \frac{1}{2}mv_i^2+m^2\sum_{i<j}u(|{\bf r}_i-{\bf r}_j|),
\label{wsm2}
\end{equation}
where $u(|{\bf r}-{\bf r}'|)$ is the  potential of interaction and $m$ the individual mass of the particles. The first term $K$ in the Hamiltonian is the kinetic energy and the second term $U$ is the potential energy. We assume that the potential of interaction decays at large distances as $r^{-\gamma}$ with $\gamma\le d$. In that case, the potential is said to be long-ranged. For such potentials, the energy $\int_0^{+\infty} u(r) r^{d-1}\, dr$ diverges at large scales implying that all the particles interact with each other and that the system displays a collective behavior.

We introduce a characteristic length $R$ and a characteristic velocity $v_m$. We define the dynamical time by $t_D=R/v_m$. For systems with long-range interactions, the potential energy scales as $U\sim N^2 m^2 u(R)$ while $K\sim N m v_m^2$. The kinetic energy and the potential energy are comparable (which is the generic situation) if $N m v_m^2\sim N^2 m^2 u(R)$. This yields $v_m^2\sim N m u(R)$. As a result, the energy scales as $E\sim N m v_m^2\sim
N^2 m^2 u(R)$ and the kinetic temperature, defined by $k_B T=m v_m^2$,
scales as $k_B T\sim N m^2 u(R)\sim E/N$. Inversely, these relations may
be used to {\it define} $R$ and $v_m$ as a function of the energy $E$
(conserved quantity in the microcanonical ensemble) or as a function
of the temperature $T$ (fixed quantity in the canonical ensemble). The proper thermodynamic limit of systems with long-range interactions
corresponds to $N\rightarrow +\infty$ in such a way that the
normalized energy $\epsilon=E/N^2m^2u(R)$ and the normalized
temperature $\eta=\beta Nm^2u(R)$ are of order
unity.  We introduce the coupling parameter $g=E_{pot}/E_{kin}=m^2 u(R)/k_B T$ where $E_{pot}\sim m^2 u(R)$ is the potential energy of two particles separated by the distance $R$ and $E_{kin}\sim k_B T$ is the thermal energy. According to the previous estimates, we have $g\sim 1/N$. Therefore, when $N\rightarrow +\infty$, we can consider a weak coupling approximation since $g\ll 1$.

It is convenient to rescale the distance by $R$, the velocity by $v_m$, the time by $t_D$, and the mass by $m$. This is equivalent to taking $R=v_m=t_D=m=1$ in the original equations. In order to satisfy the condition $N m v_m^2\sim N^2 m^2 u(R)$ the potential of interaction must scale like $u(R)\sim 1/N$. It is therefore convenient to write $u(r)=\frac{1}{N}\tilde u(r)$ with $\tilde u(R)\sim 1$ so the rescaled Hamiltonian is $H=\sum_{i=1}^N \frac{1}{2}m v_i^2+\frac{1}{N}\sum_{i<j}m^2\tilde{u}(|{\bf r}_i-{\bf r}_j|)$. This is the Kac prescription for long-range interactions \cite{kac}.  With this normalization, we have  $E\sim N$, $S\sim N$  and $T\sim 1$ in the limit $N\rightarrow +\infty$. The energy and the entropy are extensive but they remain fundamentally non-additive \cite{cdr}. The temperature is intensive. This normalization
is very convenient since the length, velocity, time and mass scales are of
order unity. Furthermore, since the coupling constant $u$ scales as $1/N$,
this immediately shows that a regime of weak coupling holds when
$N\gg 1$.

From the Liouville equation, we can derive the BBGKY hierarchy of equations for the reduced distribution functions. This hierarchy of equations can also be expressed in terms of the correlation functions (an equivalent hierarchy of equations may be obtained from the Klimontovich equation). The previous rescaling shows that only the dimensionless (coupling) parameter $g=1/N$ arises in the equations. For $N\rightarrow +\infty$, we can consider an expansion of the equations of the BBGKY hierarchy in powers of this small parameter $g=1/N\ll 1$. This corresponds to the weak coupling approximation. It is usually argued that the correlation functions scale as $P'_n\sim 1/N^{n-1}$. In particular, the two-body correlation function scales as $P'_2\sim 1/N$ and the three-body correlation function scales as $P'_3\sim 1/N^2$. This gives a systematic procedure to expand the BBGKY hierarchy in powers of the coupling parameter $g=1/N\ll 1$.

For self-gravitating systems in which the constituents interact with a potential
$u=-G/r$, we introduce the Jeans wavenumber $k_J=(4\pi G\beta m\rho)^{1/2}$ and
the gravitational pulsation $\omega_G=(4\pi G\rho)^{1/2}$, where $\rho=n m$ is
the mass density and $\beta=1/(k_B T)$ the inverse kinetic temperature. We may
use the Jeans length $\lambda_J=2\pi/k_J$ as a relevant lengthscale and the
dynamical time $t_D=2\pi/\omega_G$ as a relevant timescale. From the virial
theorem, the Jeans length gives an estimate of the system's size $R$ and the
dynamical time may be written as $t_D\sim \lambda_J/v_m\sim R/v_m$. We note that these scalings
may be obtained from the general arguments given above. We then find that the
only  dimensionless parameter in the problem is the coupling parameter
$g=1/\Lambda$ where $\Lambda=n\lambda_J^3\sim N$ gives the typical number of
particles in the Jeans sphere (i.e. in the system). Alternatively, if we define
units of length, time, velocity and mass such that $\lambda_J=t_D=v_m=m=1$,
we must take $G\sim 1/N$ for consistency. The weak coupling approximation
corresponds to $\Lambda\sim N\rightarrow +\infty$.

For Coulombian plasmas in which the constituents interact with
 a potential $u=(e^2/m^2)/r$, we introduce the Debye wavenumber $k_D=(4\pi
e^2\beta \rho/m)^{1/2}$ and the plasma pulsation $\omega_P=(4\pi
e^2\rho/m^2)^{1/2}$, where $\rho=n m$ is the mass density. We may use the Debye
length $\lambda_D=2\pi/k_D$ as a relevant lengthscale and the dynamical time
$t_D= 2\pi/\omega_P$ as a relevant timescale. The Debye length gives an estimate
of the effective range of interaction due to screening by opposite charges and
the dynamical time may be written as $t_D\sim \lambda_D/v_m$.   We then find
that the only  dimensionless parameter in the problem is the coupling parameter
$g=1/\Lambda$ where $\Lambda=n\lambda_D^3$ gives the typical number of particles
in the Debye sphere. Alternatively, if we define units of length, time, velocity
and mass such that $\lambda_D=t_D=v_m=m=1$, we must take $e^2\sim 1/\Lambda$
for consistency. The weak coupling approximation corresponds to
$\Lambda\rightarrow +\infty$.

We show that $1/\Lambda$ may indeed be interpreted as a coupling
parameter. The coupling parameter $\Gamma$ is defined as the ratio of
the interaction strength at the mean interparticle distance
$Gm^2n^{1/3}$ (resp. $e^2n^{1/3}$) to the thermal energy $k_B T$. This
leads to $\Gamma=G m^2 n^{1/3}/k_B
T=1/(n\lambda_J^3)^{2/3}=1/\Lambda^{2/3}\sim 1/N^{2/3}$
(resp. $\Gamma=e^2 n^{1/3}/k_B
T=1/(n\lambda_D^3)^{2/3}=1/\Lambda^{2/3}$).   If we define the
coupling parameter $g$ as the ratio of the interaction strength at the
Jeans (resp. Debye) length $Gm^2/\lambda_J$ (resp. $e^2/\lambda_D$)
to the thermal energy $k_B T$, we get $g=1/\Lambda$. Therefore, the
expansion of the BBGKY hierarchy in terms of the coupling parameter
$\Gamma$ or $g$ is equivalent to an expansion in terms of the inverse
of the number of particles in the Jeans sphere
$\Lambda=n\lambda_J^3\sim N$ (resp. the inverse of the number of
particles in the Debye sphere $\Lambda=n\lambda_D^3$). The weak
coupling approximation is therefore justified when $\Lambda\gg 1$.

The kinetic equations of systems with long-range interactions are well-known.
They may be derived either from the BBGKY hierarchy or from the Klimontovich
equation. For $g=1/N\rightarrow 0$, we get the Vlasov equation. This describes a
``collisionless'' evolution. In that case, the mean field approximation is
exact. At the order $g=1/N$, for spatially homogeneous
systems, we get the Lenard-Balescu equation. This is the first order correction
to the Vlasov equation due to ``collisions'' (more properly correlations)
between particles. The generalization of kinetic theory to spatially
inhomogeneous systems, using angle-action variables, is more recent
\cite{angleaction,kindetail,heyvaerts,newangleaction,aanew}. The kinetic theory
of 2D point vortices, which form a peculiar system with long-range interactions,
is also relatively new (see \cite{kinonsager} and references therein). In the
following, we summarize the basic kinetic equations of spatially homogeneous
systems with long-range interactions appropriately generalized to arbitrary
potentials of interaction (see Paper I for the details of their derivation) and
we discuss the results depending on the dimension of space. We distinguish the
evolution of the system as a whole and the relaxation of a test particle in a
bath that may be at equilibrium or out-of-equilibrium. We discuss the scaling of
the relaxation time with the number of particles.\footnote{In this paper, we
assume that the potential is non-singular at the origin. The case of pure
power-law potentials that are divergent at the origin is considered in Paper
III.} In Paper III, we apply these kinetic equations to particular systems of
physical
interest.

\section{Evolution of the system as a whole: the Lenard-Balescu equation}
\label{sec_whole}

\subsection{Spatially homogeneous systems in $d>1$}
\label{sec_ws}

Spatially homogeneous systems with long-range interactions are governed by the Lenard-Balescu equation
\begin{equation}
\frac{\partial f}{\partial t}=\pi (2\pi)^{d}m\frac{\partial}{\partial v_i}  \int d{\bf k} \, d{\bf v}'  \, k_ik_j  \frac{\hat{u}(k)^2}{|\epsilon({\bf k},{\bf k}\cdot {\bf v})|^2}\delta\lbrack {\bf k}\cdot ({\bf v}-{\bf v}')\rbrack\left (\frac{\partial}{\partial {v}_{j}}-\frac{\partial}{\partial {v'}_{j}}\right )f({\bf v},t)f({\bf v}',t),
\label{ws1}
\end{equation}
where $\hat{u}({k})$ is the Fourier transform of the potential of interaction and $\epsilon({\bf k},\omega)$ is the dielectric function
\begin{eqnarray}
\epsilon({\bf k},\omega)=1+(2\pi)^{d}\hat{u}({k})\int \frac{{\bf k}\cdot \frac{\partial f}{\partial {\bf v}}}{\omega-{\bf k}\cdot {\bf v}}\, d{\bf v}. \label{ws2}
\end{eqnarray}
The distribution function $f({\bf v},t)$ is normalized such that $\int f\, d{\bf v}=\rho$ gives the mass density.
This kinetic description assumes that the distribution function $f({\bf v},t)$ is Vlasov stable at any stage
of the dynamics.  The Lenard-Balescu equation is valid at the order $1/N$ in the limit $N\rightarrow +\infty$
(as explained in Sec. \ref{sec_scaling}, this scaling may be seen on Eq. (\ref{ws1}) by using dimensionless
variables such that $|{\bf r}|\sim 1$, $|{\bf v}|\sim 1$, $m\sim 1$, $u\sim 1/N$, and working with the one-body
distribution $P_1=f/N\sim 1$).  This corresponds to the weak coupling approximation in which three-body collisions and strong collisions may be neglected. The evolution of the system is then due exclusively to weak collisions
and collective effects (see Appendix A of Paper III).

If we neglect collective effects, which amounts to replacing $|\epsilon({\bf k},{\bf k}\cdot {\bf v})|$ by $1$, we get the Landau equation
\begin{equation}
\frac{\partial f}{\partial t}=\pi (2\pi)^{d}m\frac{\partial}{\partial v_i}  \int d{\bf k} \, d{\bf v}'  \, k_ik_j  \hat{u}(k)^2\delta\lbrack {\bf k}\cdot ({\bf v}-{\bf v}')\rbrack\left (\frac{\partial}{\partial {v}_{j}}-\frac{\partial}{\partial {v'}_{j}}\right )f({\bf v},t)f({\bf v}',t).
\label{ws3}
\end{equation}
Inversely, the Lenard-Balescu equation may be obtained from the Landau equation by replacing
the ``bare'' potential of interaction $\hat{u}({k})$ by a ``dressed'' potential of interaction
\begin{equation}
\hat{u}_{dressed}({\bf k},{\bf k}\cdot {\bf v})=\frac{\hat{u}({k})}{|\epsilon({\bf k},{\bf k}\cdot {\bf v})|}.
\label{ws4}
\end{equation}
Physically, the particles are ``dressed'' by their polarization cloud. We note that the Lenard-Balescu equation takes into account {\it dynamical} screening since the velocity of the particles explicitly appears in the dressed potential of interaction. If we replace  $|\epsilon({\bf k},{\bf k}\cdot {\bf v})|$ by  $|\epsilon({\bf k},0)|$ and approximate $f$ by the Maxwellian distribution (\ref{tb1}), we get the Debye-H\"uckel potential (see Paper I):
\begin{equation}
\hat{u}_{DH}(k)=\frac{\hat{u}(k)}{1+(2\pi)^d \hat{u}(k)\beta m\rho}.
\label{ws5}
\end{equation}
In that case, we obtain
\begin{equation}
\frac{\partial f}{\partial t}=\pi (2\pi)^{d}m\frac{\partial}{\partial v_i}  \int d{\bf k} \, d{\bf v}'  \, k_ik_j  \frac{\hat{u}(k)^2}{\left\lbrack 1+(2\pi)^d \hat{u}(k)\beta m\rho\right\rbrack^2}\delta\lbrack {\bf k}\cdot ({\bf v}-{\bf v}')\rbrack\left (\frac{\partial}{\partial {v}_{j}}-\frac{\partial}{\partial {v'}_{j}}\right )f({\bf v},t)f({\bf v}',t).
\label{ws1dh}
\end{equation}
We shall call this equation the Debye-H\"uckel approximation of the Lenard-Balescu equation.
The Debye-H\"uckel potential takes into account {\it static} screening (see
Appendix \ref{sec_dha}). Indeed, the limit $\omega={\bf k}\cdot {\bf v}\rightarrow 0$  corresponds to
the limit $t\rightarrow +\infty$ for the inverse Laplace transform of the
effective potential (\ref{ws4}). The static expression is reached when all the
transient oscillations have been damped. Alternatively, the Debye-H\"uckel
potential (\ref{ws5}) may be seen
as an approximation of Eq. (\ref{ws4}) for small velocities
$|{\bf v}|\rightarrow 0$. The Landau equation evaluated with
the Debye-H\"uckel potential will be an acceptable approximation of the
Lenard-Balescu equation whenever the large-velocity population does not play a
crucial role as compared to the bulk of the distribution. The validity of the
Debye-H\"uckel approximation of the Lenard-Balescu equation is discussed in Paper III.

For the Landau equation (\ref{ws3}) the integral over ${\bf k}$ can be performed explicitly (see, e.g., \cite{kindetail,aanew}) and we get
\begin{equation}
\frac{\partial f}{\partial t}=K_d\frac{\partial}{\partial v_i}\int d{\bf v}' \frac{w^2\delta_{ij}-w_iw_j}{w^3}\left (\frac{\partial}{\partial {v}_{j}}-\frac{\partial}{\partial {v'}_{j}}\right )f({\bf v},t)f({\bf v}',t),
\label{ws6}
\end{equation}
where ${\bf w}={\bf v}-{\bf v}'$ is the relative velocity and $K_d$ is a constant with value $K_3=8\pi^5m\int_0^{+\infty}k^3\hat{u}(k)^2\, dk$ in $d=3$ and  $K_2=8\pi^3 m\int_0^{+\infty}k^2\hat{u}(k)^2\, dk$ in $d=2$. In the Debye-H\"uckel approximation, $K_d$ is replaced by $K_d^{DH}$ defined in terms of $\hat{u}_{DH}(k)$. The Landau equation may be expressed in terms of the Rosenbluth potentials as shown in Appendix \ref{sec_rosen}.

The Lenard-Balescu  equation (\ref{ws1}) is valid at the order $1/N$ so it describes the ``collisional'' evolution of the system on a timescale $\sim N t_D$, where $t_D$ is the dynamical time.  This kinetic equation conserves the mass $M=\int f\, d{\bf v}$ and the energy $E=\int f\frac{v^2}{2}\, d{\bf v}$. It also monotonically increases the Boltzmann entropy $S=-\int \frac{f}{m}\ln \frac{f}{m}\, d{\bf v}$. More precisely, for $d>1$, we can show that $\dot S\ge 0$ and $\dot S=0$ if, and only if, $f$ is the Maxwell-Boltzmann distribution  ($H$-theorem). The collisional
evolution of the system is due to a condition of resonance between the orbits of the
particles. The condition of resonance, encapsulated in the $\delta$-function, corresponds to ${\bf k}\cdot {\bf v}'={\bf k}\cdot {\bf v}$ with ${\bf v}'\neq {\bf v}$. For $d>1$,  the condition of resonance can always be satisfied. Therefore, the only steady state of the  Lenard-Balescu equation is the Maxwell-Boltzmann distribution. This is the maximum entropy state at fixed mass and energy. Because of the H-theorem, the Lenard-Balescu  equation relaxes towards the Maxwell-Boltzmann distribution for $t\rightarrow +\infty$.  Since the collision term is valid at the order $O(1/N)$,
the relaxation time scales as
\begin{eqnarray}
t_{R}^{whole}\sim Nt_D, \qquad (d>1  \,\, {\rm homogeneous}).
 \label{ws7}
\end{eqnarray}

\subsection{Spatially homogeneous systems in $d=1$}
\label{sec_wu}

For one-dimensional systems, the Lenard-Balescu equation (\ref{ws1}) reduces to
 \begin{eqnarray}
\frac{\partial f}{\partial t}=2\pi^2 m\frac{\partial}{\partial v}  \int d{k} \, d{v}'  \, |k|  \frac{\hat{u}(k)^2}{|\epsilon({k},k v)|^2}\delta ({v}-{v}')\left (\frac{\partial}{\partial {v}}-\frac{\partial}{\partial {v'}}\right )f({v},t)f({v}',t)=0,
\label{wu1}
\end{eqnarray}
where we have used the identity $\delta(\lambda
x)=\frac{1}{|\lambda|}\delta(x)$. Therefore, the collision term
vanishes at the order $1/N$ because there is no
resonance. The kinetic equation reduces to $\partial f/\partial t=0$
so the distribution function does not evolve at all on a
timescale $\sim Nt_{D}$. To first order in $1/N$, the detailed balance
between drag and diffusion is valid not only at thermal equilibrium
but for any Vlasov stable distribution $f(v)$.  Therefore, the kinetic theory predicts no thermalization to a Maxwellian at first order in $1/N$. The Maxwellization is at least a second order effect in $1/N$, and consequently a very slow process. This result has been known for a long time in 1D plasma physics \cite{feix,kp} (in that context $N$ represents the number of particles in the Debye segment usually denoted $\Lambda=n\lambda_D$) and was rediscovered recently in the context of the HMF model \cite{bd,cvb}. This implies that, for one-dimensional homogeneous systems, the relaxation time towards
statistical equilibrium is larger than $Nt_D$. Therefore
\begin{eqnarray}
t_{R}^{whole}> N t_D, \qquad (d=1  \,\, {\rm homogeneous}).
 \label{wu2}
\end{eqnarray}
Since the relaxation process is due to more complex correlations than simply two-body collisions, we have to develop the kinetic theory at higher orders (taking  three-body, four-body,... correlation functions into account) in order to obtain the relaxation time. If the collision term does not vanish at the next order of the expansion in powers of $1/N$, the relaxation time is of order
\begin{eqnarray}
t_{R}^{whole}\sim N^2 t_D, \qquad (d=1  \,\, {\rm homogeneous}).
 \label{wu2b}
\end{eqnarray}
This quadratic scaling, conjectured in \cite{feix}, was numerically
observed in \cite{dawson,rouetfeix} for spatially homogeneous 1D plasmas. In
that case, the relaxation is caused by three-body correlations.
Similarly, for the spatially homogeneous HMF model, the above
arguments predict a relaxation time larger than $N$, scaling presumably as
$N^2$ \cite{kindetail,paper1,newangleaction}. For supercritical energies where
the system is spatially homogeneous, Campa {\it et al.} \cite{campa} report a
very long relaxation time. This could be an exponential relaxation time $\sim
e^N$, but this could also be a $N^2$ scaling with a large prefactor.

\subsection{Spatially inhomogeneous systems}
\label{sec_wi}

For spatially inhomogeneous systems, the Lenard-Balescu equation and
the Landau equation must be written with angle-action
variables
\cite{angleaction,kindetail,heyvaerts,newangleaction,aanew}.\footnote{
Angle-action variables can be introduced only when the
one-particle dynamics is integrable. This situation is encountered in specific
cases (1D systems, 3D systems with spherical symmetry) but is not generally
true. This restricts the possible applications of this formalism. Furthermore,
the Landau and Lenard-Balescu equations written with angle-action variables are
very complicated to solve numerically because we need to follow not only the
evolution of the distribution function but also the evolution of the
angle-action variables. Related to the kinetic theory, the Landau damping around
inhomogeneous states has been investigated in \cite{barre1,barre2,barre3}.} It
is then found that spatial inhomogeneity allows for additional resonances
\cite{angleaction,kindetail}. As a result, the relaxation time scales as
\begin{eqnarray}
t_{R}^{whole}\sim Nt_D, \qquad ({\rm inhomogeneous})
\label{wi0}
\end{eqnarray}
in any dimension of space (including the dimension $d=1$). This linear scaling has been observed numerically in \cite{joyce} for spatially inhomogeneous 1D self-gravitating systems, whereas for spatially homogeneous 1D plasmas  \cite{dawson,rouetfeix} the relaxation time scales as $N^2t_D$ as explained previously.

In certain cases, one can implement a local approximation and compute the collision term as if the system were spatially homogeneous. In that case, one obtains the Vlasov-Lenard-Balescu equation
\begin{equation}
\frac{\partial f}{\partial t}+{\bf v}\cdot\frac{\partial f}{\partial {\bf r}}-\nabla\Phi\cdot\frac{\partial f}{\partial {\bf v}}=\pi (2\pi)^{d}m\frac{\partial}{\partial v_i}  \int d{\bf k} \, d{\bf v}'  \, k_ik_j  \frac{\hat{u}(k)^2}{|\epsilon({\bf k},{\bf k}\cdot {\bf v})|^2}\delta\lbrack {\bf k}\cdot ({\bf v}-{\bf v}')\rbrack\left (\frac{\partial}{\partial {v}_{j}}-\frac{\partial}{\partial {v'}_{j}}\right )f({\bf r},{\bf v},t)f({\bf r},{\bf v}',t),
\label{wi1}
\end{equation}
where $\Phi({\bf r},t)=\int u(|{\bf r}-{\bf r}'|)\rho({\bf r}',t)\, d{\bf r}'$ is the self-consistent potential created by the particles. In this equation, the effects of spatial inhomogeneity are retained only in the advection term (l.h.s.) while the collision term (r.h.s.) is calculated as if the system were spatially homogeneous. If we neglect collisions, we get the Vlasov equation which is valid for $N\rightarrow +\infty$. If we neglect collective effects, replacing $|\epsilon({\bf k},{\bf k}\cdot {\bf v})|$ by $1$, we obtain the Vlasov-Landau equation
\begin{equation}
\frac{\partial f}{\partial t}+{\bf v}\cdot\frac{\partial f}{\partial {\bf r}}-\nabla\Phi\cdot\frac{\partial f}{\partial {\bf v}}=\pi (2\pi)^{d}m\frac{\partial}{\partial v_i}  \int d{\bf k} \, d{\bf v}'  \, k_ik_j  {\hat{u}(k)^2}\delta\lbrack {\bf k}\cdot ({\bf v}-{\bf v}')\rbrack\left (\frac{\partial}{\partial {v}_{j}}-\frac{\partial}{\partial {v'}_{j}}\right )f({\bf r},{\bf v},t)f({\bf r},{\bf v}',t)
\label{wi2}
\end{equation}
which may also be written as
\begin{equation}
\frac{\partial f}{\partial t}+{\bf v}\cdot\frac{\partial f}{\partial {\bf r}}-\nabla\Phi\cdot\frac{\partial f}{\partial {\bf v}}=K_d\frac{\partial}{\partial v_i}\int d{\bf v}' \frac{w^2\delta_{ij}-w_iw_j}{w^3}\left (\frac{\partial}{\partial {v}_{j}}-\frac{\partial}{\partial {v'}_{j}}\right )f({\bf r},{\bf v},t)f({\bf r},{\bf v}',t).
\label{wi3}
\end{equation}
We note that, for $d=1$, the collision term vanishes so that the kinetic equations based on the local approximation reduce to the Vlasov equation. This would imply a relaxation time $>Nt_D$. Since the rigorous treatment of spatial inhomogeneity leads to a relaxation time $\sim Nt_D$ (see above), we conclude that the local approximation is not valid in $d=1$ (see Paper III).

{\it Remark:} If a 1D system is initially spatially homogeneous but, evolving under finite $N$ effects (collisions), becomes Vlasov unstable and undergoes a dynamical phase transition towards a spatially inhomogeneous distribution, we expect that the relaxation time will scale like
\begin{eqnarray}
t_{R}^{whole}\sim N^{\delta}t_D, \qquad (d=1 {\rm \,\, dynamical \,\, phase \,\, transition})
\label{wi0b}
\end{eqnarray}
with $1\le\delta\le 2$ intermediate between Eqs. (\ref{wu2b}) and (\ref{wi0}). Such an anomalous scaling with $\delta=1.7$ has
been observed for the HMF model \cite{yamaguchi} in the situation where the system undergoes a dynamical phase transition from a non-magnetized state to a magnetized state \cite{campaall}.

\section{Relaxation of a test particle in a bath: the Fokker-Planck equation}
\label{sec_relaxtest}

We now consider the relaxation of a ``test'' particle (tagged particle) in a bath of ``field'' particles with a steady distribution function $f({\bf v})$.  We assume that the field particles are either (i) at statistical equilibrium with the Boltzmann distribution (thermal bath), in which case their distribution does not change at all with time, or (ii) in the case of one-dimensional systems, in any stable steady state of the Vlasov equation (as we have just seen, in $d=1$, this profile does not change on a timescale of order $Nt_D$). We  study how the test particle progressively acquires the distribution of the bath due to ``collisions'' with the field particles. The test particle has a stochastic motion. The evolution of the probability density $P({\bf v},t)$  that the test particle has a velocity ${\bf v}$ at time $t$  is governed by a Fokker-Planck equation involving a diffusion term and a friction term.

\subsection{General expressions of the diffusion and friction coefficients}
\label{sec_test}

The evolution of $P({\bf v},t)$ can be obtained from the
Lenard-Balescu equation (\ref{ws1}) by considering that the
distribution of the field particles is {\it fixed} (see Paper I for more details). Thus, we replace
the time dependent distribution $f({\bf v}',t)$ by the {\it static}
distribution $f({\bf v}')$ of
the bath. This procedure transforms the
integro-differential equation (\ref{ws1}) into the  differential equation\footnote{The same result can be
obtained by starting from the  Fokker-Planck
equation (\ref{test6}) and calculating the first and second moments of the increments
of velocity (see Paper I). In plasma physics, this Fokker-Planck approach, taking collective effects into
account, was first performed by Hubbard \cite{hubbard1}.}
\begin{eqnarray}
\frac{\partial P}{\partial t}=\pi (2\pi)^{d}m\frac{\partial}{\partial v_i}  \int d{\bf k} \, d{\bf v}'  \, k_ik_j   \frac{\hat{u}(k)^2}{|\epsilon({\bf k},{\bf k}\cdot {\bf v})|^2}\delta\lbrack {\bf k}\cdot ({\bf v}-{\bf v}')\rbrack\left (\frac{\partial}{\partial {v}_j}-\frac{\partial}{\partial {v'}_j}\right )P({\bf v},t)f({\bf v}'),
\label{test1}
\end{eqnarray}
where $\epsilon({\bf k},\omega)$ is the dielectric function corresponding to the fixed
distribution function $f({\bf v})$. In fact, we can understand this result in the following manner.
Equations (\ref{ws1}) and (\ref{test1}) govern the evolution of the distribution function of a
test particle (described by the coordinate ${\bf v}$) interacting with field particles (described
by the running coordinate ${\bf v}'$). In Eq. (\ref{ws1}), all the particles are equivalent so
the distribution of the field particles $f({\bf v}',t)$ changes with time exactly like the
distribution of the test particle $f({\bf v},t)$. In Eq. (\ref{test1}), the test particle and
the field particles are not equivalent since the field particles form a ``bath''. The field
particles have a steady (given) distribution $f({\bf v}')$ while the distribution of the
test particle $P({\bf v},t)$ changes with time. In the BBGKY hierarchy, this amounts
to singling out a test particle and assuming that the distribution function of the other particles is fixed.

Equation (\ref{test1}) can be written in the form of a Fokker-Planck
equation
\begin{equation}
\label{test2}
\frac{\partial P}{\partial t}=\frac{\partial}{\partial v_{i}} \left (D_{ij}\frac{\partial P}{\partial
v_{j}}- P F_i^{pol}\right )
\end{equation}
involving a diffusion tensor
\begin{eqnarray}
D_{ij}=\pi(2\pi)^d m \int d{\bf k}\, d{\bf v}' \, k_i k_j \frac{\hat{u}(k)^2}{|\epsilon({\bf k},{\bf k}\cdot {\bf v})|^2}\delta\lbrack {\bf k}\cdot ({\bf v}-{\bf v}')\rbrack f({\bf v}')
\label{test3}
\end{eqnarray}
and a friction force due to the polarization
\begin{eqnarray}
{F}_i^{pol}=\pi (2\pi)^d m\int d{\bf k}\, d{\bf v}'\, k_i k_j \frac{\hat{u}(k)^2}{|\epsilon({\bf k},{\bf k}\cdot {\bf v})|^2}\delta\lbrack {\bf k}\cdot ({\bf v}- {\bf v}')\rbrack \frac{\partial f}{\partial {v'_j}}({\bf v}').
\label{test4}
\end{eqnarray}
Using Eq. (\ref{test3}), we can easily establish the identity
\begin{eqnarray}
\label{test13a}
\frac{\partial D_{ij}}{\partial v_j}=\frac{\partial D}{\partial v_i},
\end{eqnarray}
where $D=D_{ii}$. If the velocity distribution of the field particles is isotropic, the diffusion tensor may  be written as
\begin{equation}
D_{ij}({v})=\left \lbrack D_{\|}(v)-{1\over d-1}D_{\perp}(v)\right \rbrack {v_i v_j\over v^{2}}+{1\over d-1}D_{\perp}(v)\delta_{ij},
\label{test5}
\end{equation}
where $D_{\|}(v)$ and $D_{\perp}(v)$ are the diffusion coefficients in the directions parallel and perpendicular  to the velocity of the test particle. We note that $D=D_{ii}=D_{\|}+D_{\perp}$.

Since the diffusion tensor depends on the
velocity ${\bf v}$ of the test particle, it is useful to rewrite
Eq. (\ref{test2}) in a form that is fully consistent with the general
Fokker-Planck equation \cite{risken}:
\begin{equation}
\label{test6}
\frac{\partial P}{\partial t}=\frac{\partial^{2}}{\partial v_{i}\partial v_{j}}\left (D_{ij}P\right )-\frac{\partial}{\partial v_{i}} (P F^{friction}_{i})
\end{equation}
with
\begin{equation}
\label{test7}
D_{ij}={\langle \Delta v_i \Delta v_j\rangle\over 2\Delta t}, \qquad F^{friction}_{i}={\langle \Delta v_{i}\rangle\over \Delta t}.
\end{equation}
The total friction force ${\bf F}_{friction}$ is related to the friction due to the polarization ${\bf F}_{pol}$ by
\begin{equation}
\label{test8}
F_{i}^{friction}=F_{i}^{pol}+\frac{\partial D_{ij}}{\partial v_{j}}.
\end{equation}
Substituting Eqs. (\ref{test3}) and (\ref{test4}) in Eq. (\ref{test8}), and using an integration by parts,  we find that the total friction force may be written as
\begin{eqnarray}
F_{i}^{friction}=\pi (2\pi)^d m\int d{\bf k}d{\bf v}'\, k_i k_j f({\bf v}') \left (\frac{\partial}{\partial {v}_j}-\frac{\partial}{\partial {v'}_j}\right )\delta\lbrack {\bf k}\cdot ({\bf v}-{\bf v}')\rbrack \frac{\hat{u}(k)^2}{|\epsilon({\bf k},{\bf k}\cdot {\bf v})|^2}.\qquad\qquad
\label{test9}
\end{eqnarray}
The two expressions (\ref{test2}) and (\ref{test6}) have their own interest. The expression (\ref{test6}) where the diffusion tensor is placed after the second derivative $\partial^2(DP)$ involves the total (dynamical) friction ${\bf F}_{friction}$ and the expression (\ref{test2}) where the diffusion tensor is placed between the derivatives $\partial D\partial P$ isolates the part of the friction ${\bf F}_{pol}$ due to the polarization. As shown in \cite{kalnajs,kandrup2,hb4}, the friction by  polarization ${\bf F}_{pol}$ arises from the retroaction of the field particles to the perturbation caused by the test particle. It can be obtained from a linear response theory. It represents, however, only one component of the  dynamical friction ${\bf F}_{friction}$, the other
component being $\partial_j D_{ij}$ (see Paper I).

If we neglect collective effects, the diffusion tensor and the friction force reduce to
\begin{equation}
D_{ij}=\pi(2\pi)^d m \int d{\bf k}\, d{\bf v}' \, k_i k_j \hat{u}(k)^2\delta\lbrack {\bf k}\cdot ({\bf v}-{\bf v}')\rbrack f({\bf v}'),
\label{test10}
\end{equation}
\begin{equation}
{F}_i^{pol}=\pi (2\pi)^d m\int d{\bf k}\, d{\bf v}'\, k_i k_j \hat{u}(k)^2\delta\lbrack {\bf k}\cdot ({\bf v}-{\bf v}')\rbrack \frac{\partial f}{\partial {v'_j}}({\bf v}'),
\label{test11}
\end{equation}
\begin{eqnarray}
F_{i}^{friction}=\pi (2\pi)^d m\int d{\bf k}d{\bf v}'\, k_i k_j f({\bf v}') \left (\frac{\partial}{\partial {v}_j}-\frac{\partial}{\partial {v'}_j}\right )\delta\lbrack {\bf k}\cdot ({\bf v}-{\bf v}')\rbrack \hat{u}(k)^2.\qquad\qquad
\label{test12}
\end{eqnarray}
After a series of elementary calculations \cite{landaud,kindetail}, we find that
\begin{eqnarray}
\label{test13}
 \frac{\partial D_{ij}}{\partial v_j}=F^{pol}_i.
\end{eqnarray}
Combining Eqs.  (\ref{test8}) and (\ref{test13}), we obtain
\begin{equation}
\label{test14}
{\bf F}_{friction}=2{\bf F}_{pol}
\end{equation}
so that the friction force ${\bf F}_{friction}$ is twice the friction due to the polarization ${\bf F}_{pol}$ (if the test particle has a mass $m$ and the field particles
a mass $m_f$, the factor $2$ is replaced by the ratio $(m+m_f)/m$ (see Appendix \ref{sec_ms})).

\subsection{Thermal bath: Maxwell-Boltzmann distribution}
\label{sec_tb}

We consider a thermal bath that is formed by field particles at statistical equilibrium having the isothermal (Maxwell-Boltzmann) distribution
\begin{eqnarray}
f({\bf v})= \left (\frac{\beta m}{2\pi}\right )^{d/2} \rho\, e^{-\frac{1}{2}\beta m v^2}. \label{tb1}
\end{eqnarray}
Substituting the identity ${\partial f}/{\partial {\bf v}}=-\beta m f({\bf v}){\bf v}$ in Eq. (\ref{test4}), using the $\delta$-function to replace ${\bf k}\cdot {\bf v}'$ by ${\bf k}\cdot {\bf v}$,  and comparing the resulting expression with Eqs. (\ref{test3}) and (\ref{test5}), we find that
\begin{eqnarray}
{F}_i^{pol}=-D_{ij}({\bf v})\beta m v_j=-D_{\|}(v) \beta m v_i. \label{tb4}
\end{eqnarray}
The friction coefficient $\xi=D_{\|}\beta m$ is given by an Einstein relation expressing the fluctuation-dissipation theorem. We stress that the Einstein relation is valid for the friction due to the polarization ${\bf F}_{pol}$, not for the total (dynamical) friction  ${\bf F}_{friction}$ that has a more complex expression due to the term $\partial_j D_{ij}$.  We do not have this subtlety for the usual Brownian motion for which the diffusion coefficient is constant. For a thermal bath, using
Eq. (\ref{tb4}), the Fokker-Planck equation (\ref{test2}) takes the
form
\begin{equation}
\label{tb5}\frac{\partial P}{\partial t}=\frac{\partial}{\partial v_{i}} \left \lbrack D_{ij}({\bf v})\left (\frac{\partial P}{\partial
v_{j}}+\beta m  P v_j\right )\right\rbrack,
\end{equation}
where $D_{ij}({\bf v})$ is given by Eq. (\ref{test3}) with
Eq. (\ref{tb1}). This equation is similar to
the Kramers equation in Brownian theory \cite{risken}. However, in the present case, the diffusion coefficient is
anisotropic and depends on the velocity of the test particle. For $t\rightarrow +\infty$, the distribution of the test particle relaxes towards the Maxwell-Boltzmann
distribution $P_e({\bf v})=(\beta m/2\pi)^{d/2}e^{-\beta m v^2/2}$ of the bath.  Since the Fokker-Planck equation (\ref{tb5}) is valid at the order $1/N$,  the relaxation time scales as $t_{R}^{bath}\sim N t_D$ in any dimension of
space (see Sec. \ref{sec_relaxbath}). If the distribution $P({\bf v},t)$  is
spherically symmetric, using $D_{ij}v_j=D_{\|}(v)v_i$,
we can write the Fokker-Planck equation as
\begin{equation}
\label{tb9}\frac{\partial P}{\partial t}=\frac{1}{v^{d-1}}\frac{\partial}{\partial v} \left \lbrack v^{d-1} D_{\|}({v})\left (\frac{\partial P}{\partial
v}+\beta m  P v\right )\right\rbrack
\end{equation}
or as
\begin{eqnarray}
\label{e10}
\frac{\partial P}{\partial t}={\partial\over\partial {\bf v}}\left\lbrack
D_{\|}({v})\left ({\partial P\over\partial {\bf v}}+\beta m P {\bf v}\right
)\right\rbrack.
\end{eqnarray}
It can be obtained from an {\it effective} Langevin equation
\begin{eqnarray}
\label{e11}
\frac{d{\bf v}}{dt}=-D_{\|}(v)\beta m {\bf
v}+\frac{1}{2}\frac{\partial D_{\|}}{\partial {\bf v}}+\sqrt{2D_{\|}(v)}{\bf
R}(t),
\end{eqnarray}
where ${\bf R}(t)$ is a Gaussian white noise satisfying $\langle {\bf
R}(t)\rangle ={\bf 0}$ and $\langle {R}_i(t){R}_j(t')\rangle
=\delta_{ij}\delta(t-t')$. Since the diffusion coefficient depends
on the velocity, the noise is multiplicative. The effective random force acting
on the test particle models the fluctuations of the force created
by the field particles. It can be described by a Gaussian white noise multiplied
by a
velocity-dependent factor plus a friction force equal to the
friction by polarization ${\bf F}_{pol}=-D_{\|}(v)\beta m {\bf v}$ and
a spurious drift ${\bf F}_{spurious}=(1/2)({\partial D_{\|}}/{\partial
{\bf v}})$ attributable to the multiplicative noise (we have used the
Stratonovich representation). If we neglect the velocity dependence of
the diffusion coefficient, we recover the usual Ornstein-Uhlenbeck process
\cite{risken}.

For the Maxwell-Boltzmann distribution (\ref{tb1}), the diffusion tensor (\ref{test3}) is given by (see Paper I):
\begin{equation}
\label{tb6add} D_{ij}({\bf v})=\pi(2\pi)^d\left (\frac{\beta m}{2\pi}\right )^{1/2}\rho m\int d{\bf k}\, \frac{k_i k_j}{k}\frac{\hat{u}(k)^2}{|\epsilon({\bf k},{\bf k}\cdot {\bf v})|^2}e^{-\beta m \frac{({\bf k}\cdot {\bf v})^2}{2k^2}}.
\end{equation}
Separating the integration over the modulus of ${\bf k}$ from the integration over its orientation, and
setting $\hat{\bf k}={\bf k}/k$, we obtain (see Paper I):
\begin{equation}
\label{tb6} D_{ij}({\bf v})=\int d\hat{\bf k}\, \hat{k}_{i}\hat{k}_{j}e^{-\frac{1}{2}\beta m (\hat{\bf k}\cdot {\bf v})^{2}} {\cal D}_d\left (\sqrt{\frac{\beta m}{2}}\, \hat{\bf k}\cdot {\bf v}\right ),
\end{equation}
where
\begin{equation}
\label{tb7}
{\cal D}_d(x)=\frac{1}{2(2\pi)^{d-\frac{1}{2}}}\frac{v_m^3}{n}\int_0^{+\infty}  \frac{\eta(k)^2k^d}{\left\lbrack 1-\eta(k) B(x)\right\rbrack^{2}+\eta(k)^2C(x)^{2}}\, d{k}
\end{equation}
with
\begin{equation}
\label{tb8}
\eta(k)=-(2\pi)^d\hat{u}(k)\beta m\rho, \qquad  B(x)=1-2x
e^{-x^{2}}\int_{0}^{x}e^{y^{2}}dy, \qquad  C(x)=\sqrt{\pi}x e^{-x^2}.
\end{equation}
In these expressions, $n=\rho/m$ is the numerical density of particles and $v_m=(1/\beta m)^{1/2}$ is the root mean square (r.m.s.) velocity in one direction.

For 1D systems,  the Fokker-Planck equation reduces to
\begin{equation}
\label{tb20}\frac{\partial P}{\partial t}=\frac{\partial}{\partial v} \left \lbrack D({v})\left (\frac{\partial P}{\partial
v}+\beta m  P v\right )\right\rbrack
\end{equation}
with a diffusion coefficient
\begin{equation}
\label{tb21} D({v})=2e^{-\frac{1}{2}\beta m v^{2}} {\cal D}_1\left (\sqrt{\frac{\beta m}{2}}\, {v}\right ),
\end{equation}
where
\begin{equation}
\label{tb22}
{\cal D}_1(x)=\frac{1}{2(2\pi)^{\frac{1}{2}}}\frac{v_m^3}{n}\int_0^{+\infty}  \frac{\eta(k)^2k}{\left\lbrack 1-\eta(k) B(x)\right\rbrack^{2}+\eta(k)^2C(x)^{2}}\, d{k}.
\end{equation}
The friction by polarization satisfies the Einstein relation $F_{pol}(v)=-D(v)\beta m v$.

When collective effects are neglected, the diffusion coefficient is obtained from Eqs. (\ref{tb6}) and (\ref{tb7}) by taking $B=C=0$. In that case, we obtain
\begin{equation}
\label{tb10} D_{ij}({\bf v})={\cal D}_d\int d\hat{\bf k}\, \hat{k}_{i}\hat{k}_{j}e^{-\frac{1}{2}\beta m (\hat{\bf k}\cdot {\bf v})^{2}},
\end{equation}
where
\begin{equation}
\label{tb11}
{\cal D}_d=\frac{1}{2(2\pi)^{d-\frac{1}{2}}}\frac{v_m^3}{n}\int_0^{+\infty} \eta(k)^2k^d \, d{k}.
\end{equation}
Explicit expressions of the diffusion tensor can be obtained in various dimensions of space \cite{landaud,paper1}.

In $d=3$, we have
\begin{equation}
D_{ij}({\bf v})=\left \lbrack D_{\|}(v)-\frac{1}{2} D_{\perp}(v)\right \rbrack {v_i v_j\over v^{2}}+\frac{1}{2}D_{\perp}(v)\delta_{ij}
\label{tb13}
\end{equation}
with
\begin{eqnarray}
D_{\|}(v)=2\pi^{3/2}{\cal D}_3 \frac{G\left (\sqrt{\frac{\beta m}{2}}v\right )}{\sqrt{\frac{\beta m}{2}}v},\qquad
D_{\perp}(v)=2\pi^{3/2}{\cal D}_3 \frac{{\rm erf}\left (\sqrt{\frac{\beta m}{2}}v\right )-G\left (\sqrt{\frac{\beta m}{2}}v\right )}{\sqrt{\frac{\beta m}{2}}v},
\label{tb14}
\end{eqnarray}
\begin{eqnarray}
G(w)={2\over\sqrt{\pi}}{1\over w^{2}}\int_{0}^{w}t^{2}e^{-t^{2}}dt={1\over 2w^{2}}\biggl\lbrack {\rm erf}(w)-{2w\over \sqrt{\pi}}e^{-w^{2}}\biggr\rbrack,
\label{tb15}
\end{eqnarray}
where ${\rm erf}(w)=\frac{2}{\sqrt{\pi}}\int_0^{w}e^{-u^2}\, du$ is the error function. We note the asymptotic behaviors
\begin{eqnarray}
D_{\|}(0)=\frac{4\pi{\cal D}_3}{3},\quad D_{\perp}(0)=\frac{8\pi{\cal D}_3}{3},\quad D_{\|}(v)\sim_{+\infty}\frac{\pi^{3/2}{\cal D}_3}{\left (\sqrt{\frac{\beta m}{2}}v\right )^3},\quad D_{\perp}(v)\sim_{+\infty}\frac{2\pi^{3/2}{\cal D}_3}{\sqrt{\frac{\beta m}{2}}v}.
\label{tb16}
\end{eqnarray}
They imply $D_{ij}({\bf v})\simeq D_{\|}(0)\delta_{ij}=\frac{1}{2}D_{\perp}(0)\delta_{ij}$
when $|{\bf v}|\rightarrow 0$.

In $d=2$, we have
\begin{equation}
D_{ij}({\bf v})=\left \lbrack D_{\|}(v)-D_{\perp}(v)\right \rbrack {v_i v_j\over v^{2}}+D_{\perp}(v)\delta_{ij}
\label{tb17}
\end{equation}
with
\begin{eqnarray}
D_{\|}(v)={\cal D}_2 \pi e^{-\frac{\beta m}{4}v^2}\biggl\lbrack I_{0}\biggl (\frac{\beta m}{4}v^2\biggr )- I_{1}\biggl (\frac{\beta m}{4}v^2\biggr )\biggr\rbrack, \qquad
D_{\perp}(v)= {\cal D}_2 \pi e^{-\frac{\beta m}{4}v^2}\biggl\lbrack I_{0}\biggl (\frac{\beta m}{4}v^2\biggr )+I_{1}\biggl (\frac{\beta m}{4}v^2\biggr )\biggr\rbrack,
\label{tb18}
\end{eqnarray}
where $I_n(x)$ are the modified Bessel functions.  We note the asymptotic behaviors
\begin{eqnarray}
D_{\|}(0)=\pi{\cal D}_2,\quad D_{\perp}(0)=\pi{\cal D}_2,\quad D_{\|}(v)\sim_{+\infty}\frac{\sqrt{\pi}{\cal D}_2}{\left (\sqrt{\frac{\beta m}{2}}v\right )^3},\quad D_{\perp}(v)\sim_{+\infty}\frac{2\sqrt{\pi}{\cal D}_2}{\sqrt{\frac{\beta m}{2}}v}.
\label{tb19}
\end{eqnarray}
They imply $D_{ij}({\bf v})\simeq D_{\|}(0)\delta_{ij}=D_{\perp}(0)\delta_{ij}$
when $|{\bf v}|\rightarrow 0$.

In $d=1$, we have
\begin{equation}
\label{tb23} D({v})=2{\cal D}_1 e^{-\frac{1}{2}\beta m v^{2}}.
\end{equation}

Using these analytical results, we can explicitly check that formula (\ref{test13}) is satisfied.

In the Debye-H\"uckel approximation, the previous expressions remain valid provided that ${\cal D}_d$ is replaced by
\begin{equation}
\label{etaDH} {\cal D}^{DH}_d=\frac{1}{2(2\pi)^{d-\frac{1}{2}}}\frac{v_m^3}{n}\int_0^{+\infty} \eta_{DH}(k)^2k^d \, d{k},\quad {\rm with}\quad \eta_{DH}(k)=-(2\pi)^d\hat{u}_{DH}(k)\beta m\rho=\frac{\eta(k)}{1-\eta(k)}.
\end{equation}
This amounts to taking $B=1$ and $C=0$ in Eqs. (\ref{tb6}) and (\ref{tb7}).

For spatially inhomogeneous systems, the Fokker-Planck equation must be written with angle-action variables as in \cite{angleaction,kindetail,heyvaerts,newangleaction,aanew}. When a local approximation can be implemented, the distribution $P({\bf r},{\bf v},t)$ of a test particle in a thermal bath is governed by the Fokker-Planck  equation
\begin{equation}
\label{tb24}\frac{\partial P}{\partial t}+{\bf v}\cdot \frac{\partial P}{\partial {\bf r}}-\nabla\Phi({\bf r})\cdot \frac{\partial P}{\partial {\bf v}}=\frac{\partial}{\partial v_{i}} \left \lbrack D_{ij}({\bf r},{\bf v})\left (\frac{\partial P}{\partial
v_{j}}+\beta m  P v_j\right )\right\rbrack,
\end{equation}
where $\Phi({\bf r})=\int u(|{\bf r}-{\bf r}'|)\rho({\bf r}')\, d{\bf r}'$  is
the static potential produced by the field particles at statistical equilibrium
and $D_{ij}({\bf r},{\bf v})$ is given by Eqs. (\ref{tb6})-(\ref{tb8}) where $n$
is
replaced by $n({\bf r})$. For 1D systems, the foregoing equation reduces to
\begin{equation}
\label{tb25}\frac{\partial P}{\partial t}+{v} \frac{\partial P}{\partial {x}}-\Phi'(x) \frac{\partial P}{\partial v}=\frac{\partial}{\partial v} \left \lbrack D(x,v)\left (\frac{\partial P}{\partial
v}+\beta m  P v\right )\right\rbrack,
\end{equation}
where $D(x,v)$ is given by Eqs.
(\ref{tb21})-(\ref{tb22}) where $n$
is
replaced by $n(x)$.

\subsection{Out-of-equilibrium bath in $d=1$}
\label{sec_out}

The derivation of the Fokker-Planck equation (\ref{test1}),
relying on a {\it bath approximation}, assumes that the distribution of the
field particles is ``frozen'' so that  $f({\bf v})$ does not change in time, at
least on the timescale $Nt_D$ corresponding to the relaxation time of a test
particle in a bath. This is always true for a thermal bath, corresponding to a
distribution  at statistical equilibrium (Boltzmann), because it does not change
at all. For $d>1$, due to collisions, any other distribution function relaxes
towards the Boltzmann distribution on a timescale $Nt_D$. Therefore, for $d>1$,
only the Boltzmann distribution can form a bath (the other distributions
change on a timescale $Nt_D$). The situation is different in
$d=1$ since the Lenard-Balescu collision term cancels out so that a Vlasov
stable distribution function does not change on a timescale of order $Nt_D$ (see
Sec. \ref{sec_wu}). Therefore, any Vlasov stable distribution function can be
considered as ``frozen'' on this timescale and forms a bath. We can therefore
develop a test particle approach in an out-of-equilibrium  bath characterized by
an arbitrary Vlasov stable distribution $f({v})$ that is not necessarily the
Maxwell-Boltzmann distribution of statistical equilibrium.

For 1D systems, the expressions (\ref{test3}), (\ref{test4}) and (\ref{test8}) of the diffusion coefficient and friction force take the form
\begin{equation}
\label{out1} D(v)=4\pi^{2}m f(v)\int_{0}^{+\infty} {k\hat{u}(k)^{2}\over |\epsilon(k,kv)|^{2}}\, dk,
\end{equation}
\begin{equation}
\label{out2} F_{pol}(v)=4\pi^{2}m f'(v)\int_{0}^{+\infty}  {k\hat{u}(k)^{2}\over |\epsilon(k,kv)|^{2}}\, dk,
\end{equation}
\begin{equation}
F_{friction}(v)=F_{pol}(v)+D'(v)
\label{out3}
\end{equation}
with (see Paper I):
\begin{equation}
\label{out4}
|\epsilon(k,kv)|^{2}=\left\lbrack 1-2\pi \hat{u}(k) {\cal P}\int_{-\infty}^{+\infty}\frac{f'(u)}{u-v}\, du\right\rbrack^2+4\pi^4 \hat{u}(k)^2f'(v)^2.
\end{equation}
We note the relation
\begin{equation}
\label{out8} F_{pol}(v)=D(v)\frac{d\ln f}{dv}
\end{equation}
which can be considered as a generalized Einstein relation valid for an out-of-equilibrium
bath (see also Eq. (123) in \cite{pre}). The Fokker-Planck equation (\ref{test1}) may be written in the form
\begin{equation}
\label{out9}  {\partial P\over\partial t}={\partial\over\partial v}\biggl\lbrack D(v)\biggl ({\partial P\over\partial v}-P {d\ln f\over dv}\biggr )\biggr\rbrack.
\end{equation}
Equation (\ref{out9}) is similar to a Fokker-Planck equation
describing the motion of a Brownian particle in a potential $U({v})=-\ln f({v})$ created by the field particles. The distribution function of the test particle $P(v,t)$
relaxes towards the distribution of the bath $f(v)$ for
$t\rightarrow +\infty$.  Since the Fokker-Planck equation (\ref{out9}) is valid at the order $1/N$,  the relaxation time scales as $t_{R}^{bath}\sim N t_D$. The fact that the distribution $P(v,t)$ of a test particle relaxes towards {\it any} stable distribution $f(v)$ of the field particles in $d=1$ explains why the distribution of the field particles  does not
change on a timescale $\sim Nt_{D}$. As we have said, the situation is different in $d>1$ (see further discussion in \cite{hb2,landaud}).

For an isothermal distribution (thermal bath), we recover the results of Sec. \ref{sec_tb}. For the waterbag distribution, using Eq. (B.3) of Paper I, we obtain
\begin{equation}
\label{out5}
D(v)= \frac{\rho m}{2v_m}\int_{0}^{+\infty} dk {4\pi^2\hat{u}(k)^{2}k\over \left\lbrack 1+\frac{2\pi\hat{u}(k)\rho}{v_m^2-v^2}\right\rbrack^2}
\end{equation}
for $-v_m\le v\le v_m$ and $D(v)=0$ otherwise.

When collective effects are neglected, the diffusion coefficient and the friction by polarization reduce to
\begin{equation}
\label{out6} D(v)=m f(v)\int_{0}^{+\infty}  4\pi^{2}\hat{u}(k)^{2} k\, dk,
\end{equation}
\begin{equation}
\label{out7} F_{pol}(v)=m f'(v)\int_{0}^{+\infty}  4\pi^{2} {\hat{u}(k)^{2}} k\, dk.
\end{equation}
They are respectively proportional to the distribution function $f(v)$ and to its derivative $f'(v)$.
We immediately check that $D'(v)=F_{pol}(v)$ and $F_{friction}(v)=2F_{pol}(v)$ in agreement
with Eqs. (\ref{test13}) and (\ref{test14}).

\subsection{The relaxation time}
\label{sec_relaxbath}

The relaxation time of a test particle in a bath may be
estimated by $t_R^{bath}\sim v_m^2/D\sim 1/(D\beta m)\sim 1/\xi$ where $D$ is the typical value of the diffusion coefficient. Using Eq. (\ref{tb11}) to estimate $D$, we obtain
\begin{equation}
\label{rb1}
t_{R}^{bath}\sim \frac{n}{v_m}\frac{1}{\int_0^{+\infty} \eta(k)^2 k^d\, dk},\quad {\rm or, equivalently, }\quad t_{R}^{bath}\sim \frac{v_m^3}{n m^2}\frac{1}{\int_0^{+\infty} \hat{u}(k)^2 k^d\, dk}.
\end{equation}
We can take collective effects into account in the Debye-H\"uckel approximation by simply replacing $\hat{u}(k)$ by $\hat{u}_{DH}(k)$, or $\eta(k)$ by $\eta_{DH}(k)$. Using $\eta\sim 1$, $k\sim 1/R$, $n\sim N/R^d$ and $t_D\sim R/v_m$, we get
\begin{equation}
t_{R}^{bath}\sim N t_D, \qquad (\forall d).
\label{rb2}
\end{equation}
This is the timescale controlling the relaxation of a test particle in a bath, i.e. the time needed by a
test particle to acquire the distribution of the bath. The timescale $t_{R}^{bath}$ should not be confused
with the timescale $t_R^{whole}$ controlling the relaxation of the system as a whole (see Sec. \ref{sec_whole}).
In $d>1$ they are equivalent ($t_R^{whole}\sim t_R^{bath}$) but for spatially homogeneous systems in $d=1$ they are totally different
($t_R^{whole}> t_R^{bath}$) since $t_R^{whole}>Nt_D$. In $d=1$ dimension,
the distribution $P(v,t)$ of a test particle relaxes towards the distribution of the bath $f(v)$  (that can be different from the Maxwellian as long as it is a stable steady state of the Vlasov equation) on a timescale of order $Nt_D$  while the overall distribution of the system $f(v,t)$ does not change at all on this timescale. For one dimensional plasmas, the difference in behavior of  ``distinguished'' particles that relax on a timescale $N t_D$ and the overall population that relaxes on a timescale $N^2 t_D$  is shown in \cite{feix,rouetfeix}. As a clear sign of the inequivalence of these two descriptions (evolution of the system as a whole and relaxation of a test particle in a bath), we note that the Lenard-Balescu equation (\ref{ws1}) conserves the energy while the Fokker-Planck equation (\ref{test1}) does not.

\section{A simple approximate kinetic equation which does not display any divergence for 3D Coulombian plasmas}
\label{sec_simp}

Prior to Lenard and Balescu, several authors \cite{temko,kadomtsev,tchen,ichikawa,willis} have attempted
to derive from the BBGKY hierarchy a kinetic equation which does not display any divergence in order
to improve the Landau theory. However, their treatment is approximate because, at some point of
their derivation, they replaced some correlation functions by their equilibrium values (obtained with
the Boltzmann distribution)  while Lenard and Balescu have been able to determine the exact
out-of-equilibrium correlation functions. Nevertheless, the approaches
 of \cite{temko,kadomtsev,tchen,ichikawa,willis} have the advantage of the simplicity
and they clearly show how the divergences in the Landau equation can be eliminated. In this Section, we synthesize and  generalize these different works and derive a simple approximate kinetic equation that does not display any divergence for 3D Coulombian plasmas. In the dominant approximation, it gives the same result as the Lenard-Balescu equation. We show, however, that this approximate equation breaks down in lower dimensions.

The first two equations of the BBGKY hierarchy are
\begin{eqnarray}
\label{simp1}
\frac{\partial f}{\partial t}(1)=-\frac{1}{m}\frac{\partial}{\partial {\bf v}_1}\cdot \int {\bf F}(2\rightarrow 1) g(1,2)\, d{\bf r}_2d{\bf v}_2,
\end{eqnarray}
\begin{eqnarray}
\label{simp2}
\frac{1}{2}\frac{\partial g}{\partial t}(1,2)+{\bf v}_1\cdot \frac{\partial g}{\partial {\bf r}_1}(1,2)-m\frac{\partial u_{12}}{\partial {\bf r}_1}\cdot \frac{\partial g}{\partial {\bf v}_1}(1,2)+(1\leftrightarrow 2)=m\frac{\partial u_{12}}{\partial {\bf r}_1}\cdot \frac{\partial }{\partial {\bf v}_1}f(1)f(2)\nonumber\\
+\int d{\bf r}_3d{\bf v}_3\frac{\partial u_{13}}{\partial {\bf r}_1}\cdot \frac{\partial f}{\partial {\bf v}_1}(1)g(2,3)+(1\leftrightarrow 2),
\end{eqnarray}
where $f(1)=f({\bf v}_1,t)$ is the distribution function and $g(1,2)=g(|{\bf r}_1-{\bf r}_2|,{\bf v}_1,{\bf v}_2,t)$ is the two-body correlation function. On the other hand, ${\bf F}(2\rightarrow 1)=-m\frac{\partial u_{12}}{\partial {\bf r}_1}$ is the force by unit of mass created by particle $2$ on particle $1$. These equations are valid at the order $1/N$. In particular, we have neglected the three-body correlation function which is of order $1/N^2$. This allows us to close the hierarchy. We note that the third term in Eq. (\ref{simp2}) is of order $1/N^2$ (see the scalings of Sec. \ref{sec_scaling}) but this term becomes important at small scales when $|{\bf r}_2-{\bf r}_1|\rightarrow 0$. Indeed, it is the term that accounts for strong collisions. We refer to \cite{aanew} for a derivation of these equations using the same notations as here\footnote{Actually, the equations in \cite{aanew} are more general since they take into account effects of spatial inhomogeneity. They reduce to Eqs. (\ref{simp1}) and (\ref{simp2}) for spatially homogeneous systems.} and for additional
discussion.

At statistical equilibrium, we expect the two-body correlation function to be of the form
\begin{eqnarray}
\label{simp3}
g(|{\bf r}_1-{\bf r}_2|,{\bf v}_1,{\bf v}_2)=f_M({\bf v}_1)f_M({\bf v}_2)\psi(|{\bf r}_1-{\bf r}_2|),
\end{eqnarray}
where $f_M$ is the Maxwellian distribution (\ref{tb1}). Substituting this ansatz in Eq. (\ref{simp2}), we get
\begin{eqnarray}
\label{simp4}
\frac{\partial \psi}{\partial {\bf r}_1}(1,2)\cdot {\bf v}_1+\beta m^2\frac{\partial u_{12}}{\partial {\bf r}_1}\cdot {\bf v}_1 \psi(1,2)+(1\leftrightarrow 2)=-\beta m^2\frac{\partial u_{12}}{\partial {\bf r}_1}\cdot {\bf v}_1
-\beta\rho m \int d{\bf r}_3\frac{\partial u_{13}}{\partial {\bf r}_1}\cdot {\bf v}_1 \psi(2,3)+(1\leftrightarrow 2).
\end{eqnarray}
This equation being true for any ${\bf v}_1$ and ${\bf v}_2$, we finally obtain
\begin{eqnarray}
\label{simp5}
\frac{\partial \psi}{\partial {\bf x}}(1,2)+\beta m^2\frac{\partial u}{\partial {\bf x}} \psi(1,2)=-\beta m^2\frac{\partial u}{\partial {\bf x}}
-\beta\rho m \int d{\bf r}_3\frac{\partial u_{13}}{\partial {\bf x}}\psi(2,3),
\end{eqnarray}
where ${\bf x}={\bf r}_1-{\bf r}_2$. This equation may also be derived from the YBG hierarchy starting from the microcanonical or canonical distributions \cite{hb1}.

If we neglect in Eq. (\ref{simp5}) strong collisions (second term) and collective effects (last term), we get after integration
\begin{eqnarray}
\label{simp6}
\psi({\bf x})=-\beta m^2 u({\bf x}).
\end{eqnarray}
In the case of plasmas, this expression is valid for $\lambda_L\ll |{\bf x}| \ll \lambda_D$ where $\lambda_L$ is the Landau length and $\lambda_D$ is the Debye length (see Paper III). This equation suggests interpreting $u_{eff}=-\psi/(\beta m^2)$ as an effective potential.

If we only neglect strong collisions, Eq. (\ref{simp5}) becomes
\begin{eqnarray}
\label{simp7}
\frac{\partial \psi}{\partial {\bf x}}(1,2)=-\beta m^2\frac{\partial u}{\partial {\bf x}}
-\beta\rho m \int d{\bf r}_3\frac{\partial u_{13}}{\partial {\bf x}}\psi(2,3).
\end{eqnarray}
Since the integral is a product of convolution, this equation can be easily solved in Fourier space leading to
\begin{eqnarray}
\label{simp8}
\hat{\psi}({\bf k})=\frac{-\beta m^2\hat{u}(k)}{1+\beta\rho m (2\pi)^d\hat{u}(k)}.
\end{eqnarray}
In the case of plasmas, this expression is valid for $|{\bf x}| \gg\lambda_L$. This is the proper generalization of the Debye-H\"uckel theory for an arbitrary long-range potential of interaction (the effective potential $u_{eff}=-\psi/(\beta m^2)$ is precisely the Debye potential $u_{DH}$ given by Eq. (\ref{dha4})).

If we only neglect collective effects, Eq. (\ref{simp5}) becomes
\begin{eqnarray}
\label{simp9}
\frac{\partial \psi}{\partial {\bf x}}(1,2)+\beta m^2\frac{\partial u}{\partial {\bf x}} \psi(1,2)=-\beta m^2\frac{\partial u}{\partial {\bf x}}.
\end{eqnarray}
This equation can be easily integrated in physical space leading to
\begin{eqnarray}
\label{simp10}
\psi({\bf x})=e^{-\beta m^2 u({\bf x})}-1.
\end{eqnarray}
In the case of plasmas, this expression is valid for $|{\bf x}| \ll\lambda_D$. Using Eqs. (\ref{simp3}) and (\ref{simp10}), we find that, when $|{\bf r}_1-{\bf r}_2|\rightarrow 0$, the two-body distribution function of the $N$-body problem defined by $N^2m^2P_2(1,2)=f_M(1)f_M(2)\left\lbrack 1+\psi(1,2)\right\rbrack$ is given by
\begin{eqnarray}
\label{simp11}
P_2({\bf r}_1,{\bf v}_1,{\bf r}_2,{\bf v}_2)=\frac{1}{Z_2(\beta)}e^{-\beta H_2({\bf r}_1,{\bf v}_1,{\bf r}_2,{\bf v}_2)},
\end{eqnarray}
where $H_2=\frac{1}{2}m v_1^2+\frac{1}{2}m v_1^2+m^2 u(|{\bf r}_1-{\bf r}_2|)$ is the Hamiltonian of the two-body problem. Therefore, at small scales, the two-body distribution  function is given by the canonical distribution of a pair of particles in interaction.\footnote{This result is often assumed in the literature (see, e.g., \cite{heggie,bt}). It is interesting to note that it can be derived from the BBGKY hierarchy or from the YBG hierarchy.} For 3D gravitational systems for which $u=-G/r$, we find that $\psi({\bf x})=-1+e^{G\beta m^2/|{\bf x}|}$. This correlation function accounts for
the formation of ``binaries''  when the stars come at a distance of the order of
the Landau length $\lambda_L=G\beta m^2$ (the stars attract each other and form
a pair). For one-component 3D plasmas with a neutralizing background for which
$u=e^2/(m^2 r)$, we find that $\psi({\bf x})=-1+e^{-\beta e^2/|{\bf x}|}$. This
correlation function accounts for the repulsion between two like-sign charges
when they come at a distance of the order of the Landau length $\lambda_L=\beta
e^2$ (like-sign charges repell each other). If we consider a
more realistic two-components plasma made of positive and negative charges, the
correlation between positive and negative charges is given by $\psi_{+-}({\bf
x})=-1+e^{\beta e^2/|{\bf x}|}$. This correlation function accounts for the
formation of ``atoms''  when opposite charges come at a distance of the order of
the Landau length $\lambda_L=\beta e^2$ (opposite-sign charges attract each
other and form a pair).

For 3D plasmas, Eqs. (\ref{simp6}), (\ref{simp8}) and (\ref{simp10}) may be written as
\begin{eqnarray}
\label{simp12}
\psi({\bf x})=-\frac{\beta e^2}{|{\bf x}|},\qquad \psi({\bf x})=-\frac{\beta e^2}{|{\bf x}|} e^{-|{\bf x}|/\lambda_D},\qquad \psi({\bf x})=-1+e^{-{\beta e^2}/{|{\bf x}|}}.
\end{eqnarray}
These equations, that are valid at different scales, may be unified in a single equation
\begin{eqnarray}
\label{simp13}
\psi({\bf x})={\rm exp}\left\lbrack -\frac{\beta e^2}{|{\bf x}|} e^{-|{\bf x}|/\lambda_D}\right\rbrack-1.
\end{eqnarray}
This expression was derived in \cite{nr,ti} in a different
manner.

Following \cite{temko,kadomtsev,tchen,ichikawa,willis}, we can use these equilibrium results to solve Eq. (\ref{simp2})  approximately. The idea is to replace the correlation function $g(|{\bf r}_1-{\bf r}_2|,{\bf v}_1,{\bf v}_2,t)$ in the third (strong collisions) and fifth (collective effects) terms of Eq. (\ref{simp2}) by the approximate expression
\begin{eqnarray}
\label{simp14}
g(|{\bf r}_1-{\bf r}_2|,{\bf v}_1,{\bf v}_2,t)\simeq f({\bf v}_1,t)f({\bf v}_2,t)\psi(|{\bf r}_1-{\bf r}_2|),
\end{eqnarray}
where $f({\bf v},t)$ is the out-of-equilibrium distribution function and  $\psi(|{\bf r}_1-{\bf r}_2|)$ is the {\it equilibrium} correlation function satisfying Eq. (\ref{simp5}). After simplification, we obtain
\begin{eqnarray}
\label{simp15}
\frac{1}{2}\frac{\partial g}{\partial t}(1,2)+{\bf v}_1\cdot \frac{\partial g}{\partial {\bf r}_1}(1,2)+(1\leftrightarrow 2)=m\frac{\partial u_{12}^{eff}}{\partial {\bf r}_1}\cdot \frac{\partial }{\partial {\bf v}_1}f(1)f(2)+(1\leftrightarrow 2),
\end{eqnarray}
where
\begin{eqnarray}
\label{simp16}
u_{eff}({\bf x})=-\frac{\psi({\bf x})}{\beta m^2}
\end{eqnarray}
is an effective potential determined by the equilibrium two-body correlation function. Eqs. (\ref{simp1}) and (\ref{simp15}) are now identical to those leading to the Landau equation [see, e.g., Eqs. (18) and (19) in \cite{aanew}] except that $u$ is replaced by $u_{eff}$ in Eq. (\ref{simp15}). We can therefore directly write down the kinetic equation satisfied by $f({\bf v},t)$ as
\begin{equation}
\frac{\partial f}{\partial t}=\pi (2\pi)^{d}m\frac{\partial}{\partial v_i}  \int d{\bf k} \, d{\bf v}'  \, k_ik_j  \hat{u}(k)\hat{u}_{eff}(k)\delta\lbrack {\bf k}\cdot ({\bf v}-{\bf v}')\rbrack\left (\frac{\partial}{\partial {v}_{j}}-\frac{\partial}{\partial {v'}_{j}}\right )f({\bf v},t)f({\bf v}',t).
\label{simp17}
\end{equation}
This equation can also be written as Eq. (\ref{ws6})  with $K_d$ replaced by $K_d^{eff}$ defined with $\hat{u}(k)\hat{u}_{eff}(k)$ in place of $\hat{u}(k)^2$. If we neglect strong collisions, we can use the Debye-H\"uckel expression (\ref{simp8}) of the correlation function, i.e. $\hat{u}_{eff}(k)=\hat{u}_{DH}(k)$, and we obtain
\begin{equation}
\frac{\partial f}{\partial t}=\pi (2\pi)^{d}m\frac{\partial}{\partial v_i}  \int d{\bf k} \, d{\bf v}'  \, k_ik_j  \frac{\hat{u}(k)^2}{1+\beta\rho m (2\pi)^d\hat{u}(k)}\delta\lbrack {\bf k}\cdot ({\bf v}-{\bf v}')\rbrack\left (\frac{\partial}{\partial {v}_{j}}-\frac{\partial}{\partial {v'}_{j}}\right )f({\bf v},t)f({\bf v}',t).
\label{simp18}
\end{equation}
We stress that Eq. (\ref{simp18}) is different from the Debye-H\"uckel approximation of the Lenard-Balescu equation  [see Eq. (\ref{ws1dh})] since it involves  $\hat{u}(k)\hat{u}_{DH}(k)$ instead of  $\hat{u}_{DH}(k)^2$. Actually, if we replace in the Lenard-Balescu equation (\ref{ws1}) the quantity $1/|\epsilon({\bf k},{\bf k}\cdot {\bf v})|^2$ by its value averaged over the velocity with the Maxwell distribution (\ref{tb1}), and use the identity (see Appendix C of \cite{reponselin}):
\begin{equation}
\left\langle \frac{1}{|\epsilon|^2}\right\rangle\equiv \frac{1}{\rho}\int \frac{f({\bf v})}{|\epsilon({\bf k},{\bf k}\cdot {\bf v})|^2}\, d{\bf v}=\frac{1}{\epsilon({\bf k},0)}=\frac{1}{1+\beta\rho m (2\pi)^d\hat{u}(k)},
\label{simp18add}
\end{equation}
we get the approximate kinetic equation (\ref{simp18}). This gives a clear interpretation to that equation.

In the thermal bath approach of Sec. \ref{sec_tb}, Eq. (\ref{simp18}) may be converted into the Fokker-Planck equation (\ref{tb5}) with a diffusion tensor given by Eq. (\ref{tb10}) with
\begin{equation}
{\cal D}_{d}^{eff}=\frac{1}{2(2\pi)^{d-1/2}}\frac{v_m^3}{n}\int_{0}^{+\infty}\frac{\eta(k)^2}{1-\eta(k)}k^d\, dk.
\label{simp19}
\end{equation}
For a Coulombian plasma, using $\eta(k)=-k_D^2/k^2$ (see Paper III), we get
\begin{equation}
{\cal D}_{d}^{eff}=\frac{1}{2(2\pi)^{d-1/2}}\frac{v_m^3}{n}k_D^4\int_{0}^{+\infty}\frac{k^{d-2}}{k^2+k_D^2}\, dk.
\label{simp20}
\end{equation}
In $d=3$, the integral in Eq. (\ref{simp20}) is convergent for $k\rightarrow 0$. Therefore, the approximate treatment of this Section is sufficient to suppress the divergence at large scales that appears in the Landau equation. On the other hand, the integral in Eq. (\ref{simp20}) displays a divergence at
small scales because we have neglected strong collisions in Eq. (\ref{simp18}). We can cure this divergence
heuristically by introducing a cut-off at the Landau length.\footnote{The value of the resulting integral is
\begin{equation}
\int_{0}^{k_L}\frac{k}{k^2+k_D^2}\, dk=\frac{1}{2}\ln\left (1+\Lambda^2\right ),
\label{simp20tr}
\end{equation}
where $\Lambda=k_L/k_D$.}
Had we taken into account the complete expression of the effective potential $u_{eff}({\bf r})$ this cut-off would not have been necessary since the correlation function (\ref{simp10}) accounts for a strong repulsion of like-sign charges when they come at a distance of the order of the Landau length.\footnote{For two-components plasmas and for self-gravitating systems, instead of a repulsion, we have an attraction between opposite-sign charges or between  stars. This leads to the formation of ``atoms'' $(+,-)$  or ``binary stars'' that requires a special treatment.} Indeed, the effective potential behaves like $\hat{u}_{eff}(k)\sim \hat{u}(k)/(k/k_L)$ for $k\rightarrow +\infty$ and the small scale divergence is suppressed. When strong collisions are taken into account, using
\begin{eqnarray}
\label{dha5b}
\hat{u}_{eff}({k})=\frac{\hat{u}(k)}{1+\frac{k_D^2}{k^2}+\frac{k}{k_L}},
\end{eqnarray}
Eq. (\ref{simp20}) may be replaced by
\begin{equation}
{\cal D}_{3}^{eff}=\frac{1}{2(2\pi)^{5/2}}\frac{v_m^3}{n}k_D^4\int_{0}^{+\infty}
\frac{k}{k^2+k_D^2+\frac{k^3}{k_L}}\, dk
\label{simp20b}
\end{equation}
which is convergent at small and large scales. In the dominant
approximation, we obtain
\begin{equation}
{\cal D}_{3}^{eff}=\frac{1}{2(2\pi)^{5/2}}{v_m^3}k_D\frac{\ln\Lambda}{\Lambda},
\label{simp21}
\end{equation}
where $\Lambda=\lambda_D/\lambda_L=n\lambda_D^3$ is the number of charges in the Debye sphere.
Similarly, $K_{3}^{eff}=(2\pi e^4/m^3)\ln\Lambda$. This is the same result as the one obtained
from the Lenard-Balescu equation in the dominant approximation (see, e.g., Paper III).
In $d=2$ the integral in Eq. (\ref{simp20}) does not present any divergence (implying that strong collisions are negligible) and we get
\begin{equation}
{\cal D}_{2}^{eff}=\frac{1}{4(2\pi)^{3/2}}v_m^3 k_D\frac{1}{\Lambda},
\label{simp22}
\end{equation}
where $\Lambda=n k_D^{-2}$. Similarly, $K_{2}^{eff}=\pi e^4/m^3 k_D$.
This slightly differs from the exact result obtained from the Lenard-Balescu
equation (see Paper III). However, the approximate treatment of this Section suppresses the
linear divergence at large scales present in the 2D Landau equation. Finally, in $d=1$, the
integral in Eq. (\ref{simp20}) displays a logarithmic divergence at large scales that
is absent in the Lenard-Balescu equation (see Paper III).\footnote{This divergence only occurs
in the thermal bath approach. In $d=1$, the kinetic equation (\ref{simp17}), like the Lenard-Balescu
equation, reduces to $\partial f/\partial t=0$. } However, this logarithmic divergence is
less severe than the quadratic divergence at large scales present in the 1D Landau
equation.

As a final remark, we note that we can regularize the divergence at small scales in the Lenard-Balescu equation for 3D plasmas by writing the kinetic equation (\ref{ws1})  in the form
\begin{equation}
\frac{\partial f}{\partial t}=\pi (2\pi)^{d}m\frac{\partial}{\partial v_i}  \int d{\bf k} \, d{\bf v}'  \, k_ik_j  \frac{\hat{u}(k)^2\chi(k)}{|\epsilon({\bf k},{\bf k}\cdot {\bf v})|^2}\delta\lbrack {\bf k}\cdot ({\bf v}-{\bf v}')\rbrack\left (\frac{\partial}{\partial {v}_{j}}-\frac{\partial}{\partial {v'}_{j}}\right )f({\bf v},t)f({\bf v}',t),
\label{simp23}
\end{equation}
where $\chi(k)\rightarrow 1$ for $k\rightarrow 0$ and $\chi(k)\rightarrow 0$ for $k\rightarrow +\infty$. The
function $\chi(k)$, which is related to $\hat{\psi}(k)$, takes into account the strong correlations between
the particles that appear at small scales. If we use the same arguments as those leading to Eq. (\ref{dha5b}),
we get $\chi(k)\sim k_L/k$ for $k\rightarrow +\infty$. In that case, Eq. (\ref{simp23}) is convergent
at both small and large scales. In the dominant approximation, we recover Eq. (\ref{simp21}). If we assume
that the particles are hard spheres\footnote{In general, the size of the particles is completely
negligible in plasmas and stellar systems. A situation where the size of the particles matters is the
formation of planetesimals and planets by gravitational instability \cite{stahl,exclusion}.}, we may take $\chi(k)\sim 1/(ak)$ for $k\rightarrow +\infty$ where $a$ is the size the particles. Assuming $a$ small, and considering the dominant approximation, we get Eq. (\ref{simp21}) where $\ln\Lambda$ is replaced by $\ln(\lambda_D/a)$. More generally, we should have  $\ln(\lambda_D/\lambda_{min})$ where $\lambda_{min}=\max\lbrace\lambda_L,a\rbrace$ (in the gravitational case, we obtain the same results with the Debye length replaced by the Jeans length or by
the size of the system).

\section{The effect of an external stochastic forcing}
\label{sec_stoch}

The effect of an external stochastic forcing on the evolution of a system with long-range interactions has been studied by Nardini {\it et al.} \cite{nardini1,nardini2} and, independently, by ourselves in Paper I. Our approaches are different and our results also differ at first sight. In this section, we would like to discuss further this interesting problem and  point out some connection between the two works.

Let us first consider the case of particles without interaction experiencing a friction force $-\xi {\bf v}$ with an inert medium and submitted to an external stochastic force ${\bf F}_e({\bf r},t)$. The equations of motion are
\begin{equation}
\frac{d{\bf r}_i}{dt}={\bf v}_i,\qquad \frac{d{\bf v}_i}{dt}=-\xi {\bf v}_i+{\bf F}_e({\bf r}_i(t),t).
\label{stoch1}
\end{equation}
We assume that ${\bf F}_e({\bf r},t)$ is a statistically homogeneous Gaussian white noise with zero mean and variance
\begin{equation}
\langle {F}_i^e({\bf r},t){F}_j^e({\bf r}',t')\rangle= \delta_{ij}\delta(t-t') C(|{\bf r}-{\bf r}'|).
\label{stoch11}
\end{equation}
Introducing the Fourier transform $2\pi k^2 P(k)$ of the correlation function of the noise, we have
\begin{equation}
C(|{\bf r}-{\bf r}'|)=2\pi \int  k^2 P(k) e^{i{\bf k}\cdot ({\bf r}-{\bf r}')}\, d{\bf k}.
\label{stoch11b}
\end{equation}
Equations (\ref{stoch1}) are similar to the Langevin equations describing the Brownian motion
of colloidal particles immersed in a fluid. However, in the standard approach \cite{risken}, the stochastic force
is written as $\sqrt{2D}{\bf R}_i(t)$ where ${\bf R}_i(t)$ is a Gaussian white noise
such that $\langle {\bf R}_i(t)\rangle={\bf 0}$ and
$\langle {R}^{\mu}_i(t) {R}^{\nu}_j(t')\rangle=\delta_{ij}\delta_{\mu\nu}\delta(t-t')$ acting independently on each particle. By contrast, in the present situation, ${\bf F}_e({\bf r},t)$ is a stochastic field acting coherently on all the particles. Therefore, despite some connections, the problems are fundamentally different. As in Paper I, we introduce the
discrete distribution function $f_d({\bf r},{\bf v},t)=\sum_i m \delta({\bf
r}-{\bf r}_i(t))\delta({\bf v}-{\bf v}_i(t))$ and
write the continuity equation as
\begin{equation}
\frac{\partial f_d}{\partial t}+{\bf v}\cdot\frac{\partial f_d}{\partial {\bf r}}+{\bf F}_e\cdot\frac{\partial f_d}{\partial {\bf v}}=\frac{\partial}{\partial {\bf v}}\cdot (\xi f_d {\bf v}).
\label{stoch2}
\end{equation}
Since we are dealing with a white noise, we have to specify the rules of stochastic calculus that we are using \cite{gardiner}. In writing the continuity equation (\ref{stoch2}), we have proceeded {\it as if} the force ${\bf F}_e({\bf r},t)$ were smooth, or {\it as if} it were a real noise with a finite correlation time. Therefore, we have used the Stratonovich rules of calculation.
Proceeding as in Paper I, we decompose the exact distribution function as
$f_d=f+\delta f$ where $f=\langle f_d\rangle$ is the
smooth distribution function and $\delta f$ represents the fluctuations around it. The average $\langle . \rangle$ may be viewed as a local ensemble average over different realizations at a given time. For a
spatially homogeneous system $f=f({\bf v},t)$. Substituting this decomposition
in Eq. (\ref{stoch2}), we get
\begin{equation}
\frac{\partial f}{\partial t}+\frac{\partial \delta f}{\partial t}+{\bf v}\cdot\frac{\partial \delta f}{\partial {\bf r}} +{\bf F}_e\cdot\frac{\partial f}{\partial {\bf v}}+{\bf F}_e\cdot\frac{\partial \delta f}{\partial {\bf v}}=\frac{\partial}{\partial {\bf v}}\cdot (\xi f {\bf v})+\frac{\partial}{\partial {\bf v}}\cdot (\xi \delta f {\bf v}).
\label{stoch3}
\end{equation}
Taking the local average of this equation, we obtain the equation of evolution of the smooth distribution function
\begin{equation}
\frac{\partial f}{\partial t}=-\frac{\partial}{\partial{\bf v}}\cdot \langle \delta f {\bf F}_e\rangle   +\frac{\partial}{\partial {\bf v}}\cdot (\xi f {\bf v}).
\label{stoch4}
\end{equation}
Subtracting Eq. (\ref{stoch4}) from Eq. (\ref{stoch3}) and neglecting the nonlinear terms (quasilinear approximation), we obtain the equation for the fluctuations
\begin{equation}
\frac{\partial \delta f}{\partial t}+{\bf v}\cdot\frac{\partial \delta f}{\partial {\bf r}}-\frac{\partial}{\partial {\bf v}}\cdot (\xi \delta f {\bf v})=-{\bf F}_e\cdot\frac{\partial f}{\partial {\bf v}}.
\label{stoch5}
\end{equation}
This equation may be solved formally as
\begin{equation}
\delta f({\bf r},{\bf v},t)=-\int_0^t G(t,t-\tau) {\bf F}_e({\bf r},t-\tau)\frac{\partial f}{\partial {\bf v}}({\bf v},t-\tau)\, d\tau,
\label{stoch6}
\end{equation}
where $G(t,t-\tau)$ is the Green function associated with the operator in the left hand side of Eq. (\ref{stoch5}), and we have assumed that $\delta f=0$ at $t=0$. From Eq. (\ref{stoch6}), we obtain
\begin{equation}
\langle \delta f F_i^e\rangle = -\frac{\partial}{\partial v_j}\int_0^t  \langle {F}_i^e({\bf r},t){F}_j^e({\bf r},t-\tau)\rangle f({\bf v},t-\tau)\, d\tau.
\label{stoch7}
\end{equation}
According to Eqs. (\ref{stoch11}) and (\ref{stoch11b}), the auto-correlation function of the external force is given by
\begin{equation}
\langle {F}_i^e({\bf r},t){F}_j^e({\bf r},t')\rangle=2D \delta_{ij}\delta(t-t')
\label{stoch8}
\end{equation}
with
\begin{equation}
D=\frac{C(0)}{2}=\pi \int  k^2 P(k) \, d{\bf k}.
\label{stoch12}
\end{equation}
Therefore, we have
\begin{equation}
\langle \delta f {\bf F}_e\rangle = -D \frac{\partial f}{\partial {\bf v}}.
\label{stoch9}
\end{equation}
Substituting this result in Eq. (\ref{stoch4}), we obtain the equation
\begin{equation}
\frac{\partial f}{\partial t}=\frac{\partial}{\partial {\bf v}}\cdot \left (D\frac{\partial f}{\partial {\bf v}}+\xi f {\bf v}\right ).
\label{stoch10}
\end{equation}
In order to recover the Maxwell distribution at statistical equilibrium, the
Einstein relation $D=\xi k_B T/m$ must hold between the diffusion coefficient,
the friction coefficient, and the temperature. We note that Eq. (\ref{stoch10})
coincides with the Fokker-Planck equation that is usually obtained from the
standard Langevin equations with the Gaussian white noise $\sqrt{2D}{\bf
R}_i(t)$ \cite{risken}. However, to obtain Eq. (\ref{stoch10}), we have
not used the usual formalism of Fokker-Planck equations (Kramers-Moyal expansion). We have
rather used a method inspired by fluctuating hydrodynamics \cite{llif,hb5}.

In Paper I, we have tried to generalize this approach to the case where the particles interact via long-range forces. In that case, we must add the force $-\nabla\Phi_d({\bf r}_i(t),t)=-\int \nabla u(|{\bf r}_i(t)-{\bf r}'|) f_d({\bf r}',{\bf v}',t)\, d{\bf r}'d{\bf v}'$ produced by the other particles in the equations of motion (\ref{stoch1}). The continuity equation then becomes
\begin{equation}
\frac{\partial f_d}{\partial t}+{\bf v}\cdot\frac{\partial f_d}{\partial {\bf r}}-\nabla\Phi_d\cdot\frac{\partial f_d}{\partial {\bf v}}+{\bf F}_e\cdot\frac{\partial f_d}{\partial {\bf v}}=\frac{\partial}{\partial {\bf v}}\cdot (\xi f_d {\bf v}).
\label{stoch13}
\end{equation}
For $\xi=0$ and ${\bf F}_e={\bf 0}$, it reduces to the Klimontovich equation. Within the quasilinear approximation, we obtain the closed coupled equations
\begin{equation}
\frac{\partial f}{\partial t}=\frac{\partial}{\partial{\bf v}}\cdot \langle \delta f \nabla\delta\Phi\rangle-\frac{\partial}{\partial{\bf v}}\cdot \langle \delta f {\bf F}_e\rangle  +\frac{\partial}{\partial {\bf v}}\cdot (\xi f {\bf v})
\label{stoch14}
\end{equation}
and
\begin{equation}
\frac{\partial \delta f}{\partial t}+{\bf v}\cdot\frac{\partial \delta f}{\partial {\bf r}}=\nabla\delta\Phi\cdot\frac{\partial f}{\partial {\bf v}}-{\bf F}_e\cdot\frac{\partial f}{\partial {\bf v}},
\label{stoch15}
\end{equation}
where we have assumed that the friction is weak ($\xi\ll 1$) so that it can be neglected in Eq. (\ref{stoch15}). From these equations, we have obtained in Paper I a nonlinear Fokker-Planck equation of the form
\begin{equation}
\frac{\partial f}{\partial t}=\frac{\partial}{\partial {v}_i} \left (D_{ij}[f]\frac{\partial f}{\partial {v}_j}+\xi f {v}_i\right )
\label{stoch16}
\end{equation}
with a diffusion tensor\footnote{When considering the effect of the external stochastic force,  the Laplace transforms in Sec. 6 of Paper I must in fact be replaced by Fourier transforms. As a result, Eqs. (141) and (C.4) of Paper I must be multiplied by $(2\pi)^2$.}
\begin{equation}
D_{ij}[f]=\pi\int {k}_i {k}_j \frac{P(k)}{|\epsilon({\bf k},{\bf k}\cdot {\bf v})|^2}\, d{\bf k}.
\label{stoch17}
\end{equation}
In the non-interacting case, we have $\epsilon({\bf k},{\bf k}\cdot {\bf v})=1$, and we recover Eq. (\ref{stoch12}) as it should.

However, as discussed by Gardiner \cite{gardiner} and others, differential equations which include a white noise as a driving term have to be handled with great care. In particular, the Stratonovich stochastic calculus can be very tricky because, in that case, the noise does not satisfy the non-anticipating property. As a result, it is almost impossible to carry out rigorous proofs with the Stratonovich  calculus and it is more convenient to use the Ito calculus. In the present case, the two interpretations (Ito and Stratonovich) should lead to the same results because, although the noise is multiplicative (it depends on the positions), it only acts on the velocities.\footnote{I am grateful to the referee for these comments.}

If we use the Ito formula in the spirit of \cite{dean,hb5}, instead of the continuity equation (\ref{stoch13}), we get an equation of the form
\begin{equation}
\frac{\partial f_d}{\partial t}+{\bf v}\cdot\frac{\partial f_d}{\partial {\bf r}}-\nabla\Phi_d\cdot\frac{\partial f_d}{\partial {\bf v}}+{\bf F}_e\cdot\frac{\partial f_d}{\partial {\bf v}}=\frac{\partial}{\partial {\bf v}}\cdot \left (D\frac{\partial f_d}{\partial {\bf v}}+\xi f_d {\bf v}\right ),
\label{stoch18}
\end{equation}
where $D$ is given by Eq. (\ref{stoch12}). Then,  Eq. (\ref{stoch14}) is replaced by
\begin{equation}
\frac{\partial f}{\partial t}=\frac{\partial}{\partial{\bf v}}\cdot \langle \delta f \nabla\delta\Phi\rangle+\frac{\partial}{\partial {\bf v}}\cdot \left (D\frac{\partial f}{\partial {\bf v}}+\xi f {\bf v}\right )
\label{stoch19}
\end{equation}
while, within the quasilinear approximation, Eq. (\ref{stoch15})
remains unchanged in the limit of weak frictions $\xi\ll 1$. In order to obtain
Eq. (\ref{stoch19}), we have treated the term $\eta\equiv
\nabla_{\bf v}\cdot (f_d{\bf F}_e)$ in Eq. (\ref{stoch18}) as a noise with zero
mean $\langle \eta\rangle=0$ (in the same spirit as in \cite{dean,hb5}) and,
consequently, we have not written the  term  $\langle\delta f {\bf F}_e\rangle$
in Eq. (\ref{stoch19}). In this way, in the absence of interaction, Eq.
(\ref{stoch19}) reduces to Eq. (\ref{stoch10}) as it should. On the other hand,
the term $\langle \delta f \nabla\delta\Phi\rangle$ coming from the long-range interaction between the particles has been explicitly computed in
Appendix C of Paper I, leading to Eq. (I-C.4). Substituting
this result in Eq. (\ref{stoch19}) we obtain a nonlinear Fokker-Planck equation
of the form
\begin{equation}
\frac{\partial f}{\partial t}=\frac{\partial}{\partial {v}_i}\left (D_{ij}^{I}[f]\frac{\partial f}{\partial {v}_j}\right )+\frac{\partial}{\partial {\bf v}}\cdot \left (D\frac{\partial f}{\partial {\bf v}}+\xi f {\bf v}\right )
\label{stoch20}
\end{equation}
with a diffusion tensor
\begin{equation}
D_{ij}^I[f]=\pi (2\pi)^d\int d{\bf k}\, k_i k_j \hat{u}(k){\cal P}\int d{\bf v}'\frac{{\bf k}\cdot \frac{\partial f}{\partial {\bf v}'}}{{\bf k}\cdot ({\bf v}'-{\bf v})}\left (\frac{1}{|\epsilon({\bf k},{\bf k}\cdot {\bf v}')|^2}+\frac{1}{|\epsilon({\bf k},{\bf k}\cdot {\bf v})|^2}\right )P(k) .
\label{stoch21}
\end{equation}
This equation coincides with the result obtained by Nardini
{\it et al.}  \cite{nardini1,nardini2} using a different approach (we can make
contact with their notations by setting $c_k=2\pi k^2 P(k)$). This equation has
been extensively tested against numerical simulations of the $N$-body system and
analytical arguments in \cite{nardini2}. The noise $\eta$ found above may
account for the fluctuations and for the dynamical phase transitions
observed by these authors in \cite{nardini2}.

Let us comment on the differences between Eqs.
(\ref{stoch16})-(\ref{stoch17}) and Eqs. (\ref{stoch20})-(\ref{stoch21}). To
obtain Eqs. (\ref{stoch20})-(\ref{stoch21}), we have used the Ito interpretation in which the noise has the property to be non-anticipating implying $\langle f_d {\bf F}_e\rangle_{I}=\langle \delta f {\bf F}_e\rangle_{I}={\bf 0}$.  As a result, only the diffusion tensor $D_{ij}^I[f]$ associated with $\langle \delta f \nabla\delta\Phi\rangle$ appears
in addition to the normal diffusion $D$. By contrast, to obtain Eqs.
(\ref{stoch16})-(\ref{stoch17}), we have used the Stratonovich interpretation
in which  the noise does not satisfy the non-anticipating property so that
$\langle f_d {\bf F}_e\rangle_{S}= \langle
\delta f {\bf F}_e\rangle_{S}\neq {\bf 0}$. In that case, both the diffusion tensor $D_{ij}^I[f]$ associated
with $\langle \delta f \nabla\delta\Phi\rangle$ and the diffusion tensor
$D_{ij}^{II}[f]$ associated with $\langle \delta f {\bf F}_e\rangle_{S}$  appear and
their sum simplifies to give the diffusion tensor $D_{ij}[f]$ (see Paper I).
If we argue that the Ito and the Stratonovich interpretations
should lead to the same result (see above),
this implies that there is a mistake in the derivation of Eqs. (\ref{stoch16})-(\ref{stoch17}). We note that Eq. (\ref{stoch16}) may be rewritten as
\begin{equation}
\frac{\partial f}{\partial t}=\frac{\partial}{\partial {v}_i}\left (D_{ij}^{I}[f]\frac{\partial f}{\partial {v}_j}\right )+\frac{\partial}{\partial {v}_i} \left (D_{ij}^{II}[f]\frac{\partial f}{\partial {v}_j}+\xi f {v}_i\right ).
\label{stoch16bis}
\end{equation}
Under this form, we see that the two equations (\ref{stoch20}) and (\ref{stoch16bis}) are relatively close. They both include the diffusion tensor $D^{I}_{ij}[f]$, whose expression is given by Eq. (\ref{stoch21}),  coming from the long-range interaction $\langle \delta f \nabla\delta\Phi\rangle$. They differ only by the fact that the bare diffusion coefficient $D\delta_{ij}$ in Eq. (\ref{stoch20}) is replaced by a more complicated diffusion coefficient $D_{ij}^{II}[f]$ in Eq. (\ref{stoch16bis}) affected by the long-range interaction (it reduces to $D\delta_{ij}$ in the absence of interaction).  Therefore, Eq.  (\ref{stoch20}) obtained with the Ito interpretation would be obtained with the Stratonovich interpretation if the diffusion tensor associated with $\langle \delta f {\bf F}_e\rangle_{S}$ were equal to $D \delta_{ij}$ (as in the absence of interaction) instead of $D_{ij}^{II}[f]$. This may be related to a subtlety in the Stratonovich calculus \cite{gardiner}.  However, we leave open the possibility that the averages are not exactly equivalent in the two approaches. Eqs. (\ref{stoch16})-(\ref{stoch17}) may correspond to using ensemble averages at  a given time $t$ while Eqs. (\ref{stoch20})-(\ref{stoch21}) may correspond to using noise averages over a small interval of time $\Delta t$.  Therefore, the difference between Eqs.  (\ref{stoch16})-(\ref{stoch17}) and
Eqs. (\ref{stoch20})-(\ref{stoch21}) may be due to a different definition of the
averages. We note that, in the absence of interaction, Eqs.
(\ref{stoch16})-(\ref{stoch17}) and Eqs. (\ref{stoch20})-(\ref{stoch21}) coincide
(they both reduce to the ordinary Fokker-Planck equation  (\ref{stoch10}) with the diffusion coefficient (\ref{stoch12})) while this is no longer true in the presence of long-range interactions. It is known that ensemble averages and time averages may give different results for long-range systems.

If we apply Eqs. (\ref{stoch16})-(\ref{stoch17}) and Eqs.
(\ref{stoch20})-(\ref{stoch21}) to the HMF model where the potential of
interaction $u=1-\cos(\theta-\theta')$ is restricted to the first Fourier modes
$n=\pm 1$, so that
$\hat{u}_n=\frac{1}{2}(2\delta_{n,0}-\delta_{n,1}-\delta_{n,-1})$, we get a
nonlinear Fokker-Planck equation of the form
\begin{equation}
\frac{\partial f}{\partial t}=\frac{\partial}{\partial {v}} \left (D[f]\frac{\partial f}{\partial {v}}+\xi f {v}\right )
\label{ass1}
\end{equation}
with a diffusion coefficient
\begin{equation}
D[f]=D-c_1\left (1-\frac{1}{|\epsilon(1,v)|^2}\right )
\label{ass2}
\end{equation}
corresponding to Eq. (\ref{stoch17}), or with a diffusion coefficient
\begin{equation}
D[f]=D+2\pi c_1 \hat{u}_1{\cal P}\int d{v}'\frac{\frac{\partial f}{\partial {v}'}}{{v}'-{v}}\left (\frac{1}{|\epsilon(1,{v}')|^2}+\frac{1}{|\epsilon(1,{v})|^2}\right )
\label{ass3}
\end{equation}
corresponding to Eq. (\ref{stoch21}), where $D$ is given by Eq. (\ref{stoch12}) yielding $D={C(0)}/{2}=c_0/2+\sum_{n=1}^{+\infty}c_n$. These diffusion coefficients clearly differ\footnote{An
important difference is that Eqs. (\ref{ass1}) and (\ref{ass3}) satisfy the
energy balance $\langle de/dt\rangle+2\xi\langle \kappa\rangle =D$ (where $e$ is the energy density and $\kappa$ is the kinetic energy density) derived in \cite{nardini2}, contrary to Eqs. (\ref{ass1}) and
(\ref{ass2}). In order to obtain this energy balance, the authors of \cite{nardini2} use the fact that $\langle {v}_i(t)\cdot {F}_e({\theta}_i(t),t)\rangle_I=0$. This is valid with the Ito interpretation in which the noise is non-anticipating but this is not true with the Stratonovich interpretation. Again, if we argue that the two interpretations should lead to the same results, the Stratonovich interpretation has to give the same energy balance as the Ito's one. This requires $\langle {v}_i(t)\cdot {F}_e({\theta}_i(t),t)\rangle_S=D$ (as in the absence of interaction). This would then imply that Eqs. (\ref{ass1}) and
(\ref{ass2}) are incorrect since they do not satisfy the energy balance $\langle de/dt\rangle+2\xi\langle \kappa\rangle =D$. However, as discussed above, the way to treat the noise and define the averages may be different in the two approaches so that  the energy balance derived in \cite{nardini2} may apply to Eqs. (\ref{ass1}) and (\ref{ass3}) but not to Eqs. (\ref{ass1}) and (\ref{ass2}) for which the energy balance may have a more complicated expression.} but they have some common
properties. For example, in the absence of interaction, they reduce to Eq.
(\ref{stoch12}) leading to the linear Fokker-Planck equation (\ref{stoch10}).
Equation (\ref{ass1}) with the diffusion coefficient (\ref{ass2}) or
(\ref{ass3}) also reduces  to the linear Fokker-Planck equation (\ref{stoch10})
if the forcing does not act on the modes $n=\pm 1$ involved in the potential, so
that $c_1=0$. By contrast, if the forcing ``excites'' the modes  $n=\pm 1$
involved in the potential, so that $c_1\neq 0$, Eq. (\ref{ass1}) with the
diffusion coefficient (\ref{ass2}) or (\ref{ass3}) becomes a nonlinear
Fokker-Planck equation.

In summary:

(i) We have shown that, in the case of particles without
interaction, the approach developed in Paper I returns the standard
Fokker-Planck equation (\ref{stoch10}). In that case, the calculations can be
performed in physical space which enlightens the basic physics. This equation is
obtained by an original approach related to fluctuating hydrodynamics without
using a Kramers-Moyal expansion or the other methods of Brownian theory and
stochastic processes. Using the same approach but now taking into account
the interactions between the particles, we have obtained in Paper I
the nonlinear Fokker-Planck equation (\ref{stoch16})-(\ref{stoch17}).
It reduces to Eq. (\ref{stoch10}) with Eq. (\ref{stoch12}) in the absence of interaction.

(ii) We have shown a relationship between our approach and the
results of Nardini {\it et al.} \cite{nardini1,nardini2}. Actually, Eq.
(\ref{stoch21}) corresponds to the diffusion tensor
$D_{ij}^{I}[f]$ associated with the long-range interaction term
$\langle \delta f\delta\Phi\rangle$ calculated in Paper I (see Eq. (I-C.4)). The
difference between Eqs.  (\ref{stoch16})-(\ref{stoch17}) and
Eqs. (\ref{stoch20})-(\ref{stoch21}) comes
therefore from the evaluation of the term  $\langle \delta f {\bf F}_e\rangle$. In the
Ito interpretation, this term is zero while this is no more true in the Stratonovich interpretation. In Paper I, we have suggested that this term gives rise to a diffusion tensor $D_{ij}^{II}[f]$ leading to
Eq.  (\ref{stoch16bis}) which differs from
Eq. (\ref{stoch20}). Alternatively, if we assume that this term leads to a diffusion tensor $\delta_{ij}D$, as in the absence of interaction, we recover Eq. (\ref{stoch20}).

\section{Conclusion}
\label{sec_conclusion}

In the introduction of this paper, we have proposed a short historic concerning the development of kinetic theory in plasma
physics. We have mentioned important contributions that are not well-known. In particular, we have noted
that, prior to the classical works of Lenard and Balescu, some authors derived an approximate kinetic equation
that solves the divergences at small and large scales appearing in the Landau equation. In the main part
of the paper, we have developed
these works further and we have shown that this approximate kinetic equation may be obtained from the
Lenard-Balescu equation by replacing the quantity $1/|\epsilon({\bf k},{\bf k}\cdot {\bf v})|^2$ by its
average $\langle 1/|\epsilon|^2|\rangle$ over the velocities computed with the Maxwell distribution. We have
also recapitulated the main kinetic equations of plasma physics and we have extended them
to arbitrary long-range potentials of interaction in different dimensions of space. We have discussed
the scaling of the relaxation time with the number of particles
and with the dimension of space. We have also discussed the effect of a
stochastic forcing
on the evolution of systems with long-range interactions. In Paper III,
we apply these results to specific systems with long-range interactions such as
systems with power-law potentials, plasmas
and stellar systems in arbitrary dimensions of space, and the HMF model with
attractive and repulsive interactions.

\appendix

\section{The Debye-H\"uckel theory for a general long-range potential of interaction}
\label{sec_dha}

In this Appendix we provide the proper generalization of the Debye-H\"uckel theory for an arbitrary long-range potential of interaction. We wish to determine the effective interaction between particles at statistical equilibrium taking  collective effects into account. A single (test) particle at ${\bf r}_P={\bf 0}$ produces a ``naked'' potential $\Phi_0({\bf r})=mu({\bf r})$. This potential modifies the distribution of the other (field) particles. The resulting change of density $\tilde{\rho}({\bf r})$ in turn produces an extra-potential which adds to the original one $\Phi_0({\bf r})$. The effective, or ``dressed'', potential  created by the test particle is therefore the solution of the self-consistency equation
\begin{eqnarray}
\label{dha1}
\Phi_{eff}({\bf r})=\int u({\bf r}-{\bf r}')\left\lbrack \tilde{\rho}({\bf r}')+m\delta({\bf r}')\right \rbrack\, d{\bf r}'.
\end{eqnarray}
The variation of the distribution of the field particles is given by the Boltzmann statistics
\begin{eqnarray}
\label{dha2}
\tilde\rho({\bf r})=\rho e^{-\beta m\Phi_{eff}({\bf r})}-\rho,
\end{eqnarray}
where $\rho$ is the uniform distribution of the particles. In the weak coupling
limit $\beta m\Phi_{eff}\ll 1$, we can expand the exponential in Eq.
(\ref{dha2}) so that $\tilde\rho\simeq -\beta m\rho\Phi_{eff}$. Substituting
this result in Eq. (\ref{dha1}) we obtain an integro-differential equation
\begin{eqnarray}
\label{dha3}
\Phi_{eff}({\bf r})=\int u({\bf r}-{\bf r}')\left\lbrack -\beta m\rho \Phi_{eff}({\bf r}')+m\delta({\bf r}')\right \rbrack\, d{\bf r}'
\end{eqnarray}
determining the effective potential. Since the integral is a product of convolution, this equation can be solved easily in Fourier space. Writing $\Phi_{eff}({\bf r})=m u_{DH}({\bf r})$, we obtain
\begin{eqnarray}
\label{dha4}
\hat{u}_{DH}({k})=\frac{\hat{u}(k)}{1+(2\pi)^d \hat{u}(k)\beta m\rho}.
\end{eqnarray}
Specializing this expression to 3D plasmas, we recover the Debye-H\"uckel potential
\begin{eqnarray}
\label{dha5}
(2\pi)^d\hat{u}_{DH}({k})=\frac{(2\pi)^d\hat{u}(k)}{1+\frac{k_D^2}{k^2}}=\frac{S_d e^2}{m^2}\frac{1}{k^2+k_D^2}.
\end{eqnarray}
In physical space, we have $u_{DH}=(e^2/m^2)r^{-1}e^{-k_D r}$ where $k_D=(4\pi e^2\beta\rho/m)^{1/2}$ is the Debye wavenumber. An equivalent result, expressed in terms of the two-body correlation function,  can  be derived from the YBG hierarchy \cite{hb1} or from the
BBGKY hierarchy (see Sec. \ref{sec_simp}). The generalization (\ref{dha4}) allows us to consider other potentials than just the Coulombian potential. Some explicit examples are given in \cite{hb1}.

\section{The Rosenbluth potentials in $d$ dimensions}
\label{sec_rosen}

In this Appendix, we show that the Landau equation may be expressed in terms of the Rosenbluth
potentials \cite{rosen}. Our results are valid in $d$ dimensions. The Landau equation (\ref{ws6})  may be written as
\begin{eqnarray}
\label{rosen0} \frac{\partial f}{\partial t}=\frac{\partial}{\partial
v_i}\int  K_{ij}\left (f'\frac{\partial
f}{\partial v_{j}}-f\frac{\partial
f'}{\partial v^{'}_j}\right )\, d{\bf v}',\quad {\rm with}\quad K_{ij}=K_d \frac{w^2\delta_{ij}-w_i w_j}{w^3},
\end{eqnarray}
where $f=f({\bf v},t)$, $f'=f({\bf v}',t)$, and ${\bf w}={\bf v}-{\bf v}'$. It can be put in the Fokker-Planck form
\begin{eqnarray}
\label{rosen12}
\frac{\partial f}{\partial t}=\frac{\partial}{\partial v_i}\left (D_{ij}\frac{\partial f}{\partial v_{j}}-f F^{pol}_i\right ),\quad {\rm or}\quad  \frac{\partial f}{\partial t}=\frac{\partial^{2}}{\partial v_{i}\partial v_{j}}\left (D_{ij}f\right )-\frac{\partial}{\partial v_{i}} (f F^{friction}_{i}),
\end{eqnarray}
where the diffusion and friction coefficients are given by
\begin{equation}
\label{rosen1} D_{ij}=\int K_{ij} f' \, d{\bf v}'=K_d\int f' \frac{w^2\delta_{ij}-w_{i}w_{j}}{w^3}\, d{\bf v}',
\end{equation}
\begin{eqnarray}
{F}_{i}^{friction}=2{F}_{i}^{pol}=2\int K_{ij} \frac{\partial f'}{\partial v^{'}_{j}} \, d{\bf v}'
=2\int \frac{\partial K_{ij}}{\partial v_{j}} f' \, d{\bf v}'=-2(d-1)K_d\int f' \frac{w_i}{w^3}\, d{\bf v}'.
\label{rosen2}
\end{eqnarray}
Using the identities
\begin{eqnarray}
\label{rosen3}
K_{ij}=K_d\frac{\partial^{2}w}{\partial v_i\partial v_j},\qquad {\rm and}\qquad \frac{\partial K_{ij}}{\partial v_j}=-(d-1)K_d\frac{w_{i}}{w^{3}}=(d-1)K_d \frac{\partial}{\partial v_{i}}\left (\frac{1}{w}\right),
\end{eqnarray}
the coefficients of diffusion and friction may be rewritten as
\begin{eqnarray}
\label{rosen5}
D_{ij}=K_d \frac{\partial^{2}g}{\partial v_{i}\partial v_{j}} ({\bf v},t),\qquad {\bf F}_{friction}=2{\bf F}_{pol}=2(d-1)K_d \frac{\partial h}{\partial {\bf v}} ({\bf v},t),
\end{eqnarray}
where
\begin{eqnarray}
\label{rosen7}
g({\bf v},t)=\int f({\bf v}',t)|{\bf v}-{\bf v}'|d{\bf v}',\quad h({\bf v},t)=\int \frac{f({\bf v}',t)}{|{\bf v}-{\bf v}'|}d{\bf v}'
\end{eqnarray}
are the so-called Rosenbluth potentials \cite{rosen}. In terms of these potentials, the Landau equation may be rewritten as
\begin{eqnarray}
\label{rosen8}
\frac{\partial f}{\partial t}=K_d\frac{\partial}{\partial v_i}\left\lbrack \frac{\partial^2g}{\partial v_i\partial v_j}\frac{\partial f}{\partial v_j}-(d-1)f\frac{\partial h}{\partial v_i}\right\rbrack,\quad {\rm or}\quad \frac{\partial f}{\partial t}=K_d\left\lbrack \frac{\partial^{2}}{\partial v_{i}\partial v_{j}}\left ( \frac{\partial^2g}{\partial v_i\partial v_j}f\right )-2(d-1)\frac{\partial}{\partial v_{i}} \left (f\frac{\partial h}{\partial v_i}\right )\right\rbrack.
\end{eqnarray}

For an isotropic distribution function $f=f(v,t)$, implying $g=g(v,t)$ and $h=h(v,t)$, the diffusion tensor (\ref{rosen5}-a) may be written as
\begin{equation}
\label{rosen9}
D_{ij}=\left (D_{\|}-\frac{1}{d-1}D_{\perp}\right )\frac{v_i v_j}{v^{2}}+\frac{1}{d-1}D_{\perp}\delta_{ij}
\end{equation}
with
\begin{eqnarray}
\label{rosen10}
D_{\|}=K_d\frac{\partial^{2}g}{\partial v^{2}},\qquad D_{\perp}=(d-1)K_d\frac{1}{v}\frac{\partial g}{\partial v}.
\end{eqnarray}
On the other hand, the friction force (\ref{rosen5}-b) may
be written as
\begin{eqnarray}
\label{rosen11}
{\bf F}_{friction}=2{\bf F}_{pol}=2(d-1)K_d\frac{1}{v}\frac{\partial h}{\partial v}{\bf v}.
\end{eqnarray}
Using $D_{ij}v_{j}=D_{\|}v_{i}$, the Landau equation (\ref{rosen12}) becomes
\begin{eqnarray}
\label{rosen15}
\frac{\partial f}{\partial t}=K_d\frac{1}{v^{d-1}}\frac{\partial}{\partial v}\left\lbrack v^{d-1}\left (\frac{\partial^2 g}{\partial v^2}\frac{\partial f}{\partial v}-(d-1) f\frac{\partial h}{\partial v}\right )\right\rbrack.
\end{eqnarray}
Explicit expressions of $g(v,t)$ and $h(v,t)$ in terms of $f(v,t)$ are given in \cite{bt} in $d=3$ (see also \cite{aanew}). The Landau equation (\ref{rosen15})  describes the evolution of the system as a whole. This Landau-type equation, corresponding to a microcanonical description in which the energy is conserved, was used by King \cite{kingL} to study the dynamics of globular clusters. If we replace $f({v},t)$ by $P({v},t)$ and  $f({v}',t)$ by the Maxwellian distribution $f({v}')$ given by Eq. (\ref{tb1}), we obtain the Fokker-Planck equation (\ref{tb9}) describing the relaxation of a test particle in a bath of field particles at statistical equilibrium (thermal bath). In that case, we recover the results of Sec. \ref{sec_tb}. This Fokker-Planck-type equation, corresponding to a canonical description in which the temperature is fixed, was used by Chandrasekhar \cite{chandra1,chandra2} to determine the evaporation rate of globular clusters (see \cite{lc} for
a more detailed discussion of the difference between the approaches of Chandrasekhar and King). Eq. (\ref{rosen15}) may be used to obtain a more general Fokker-Planck equation. Indeed, if we replace $f({v},t)$ by $P({v},t)$ and  $f({v}',t)$ by  {\it any} isotropic distribution $f({v}')$, we obtain a Fokker-Planck equation describing the relaxation of a test particle in an out-of-equilibrium bath of field particles.\footnote{Actually, this more general approach is not self-consistent in $d>1$ since, as already explained, the distribution $f({v})$ is not steady on the timescale $t_R\sim N t_D$ over which the test particle relaxes, unless $f(v)$ is the Maxwell distribution \cite{landaud,hb2}.}  We recall, however, that the results of this Appendix assume that the system is spatially homogeneous (or that a local approximation can be implemented) and neglects collective effects.

\section{Multi-species systems}
\label{sec_ms}

It is straightforward to generalize the kinetic theory to several species of particles. The Lenard-Balescu equation
(\ref{ws1}) is replaced by
\begin{equation}
\frac{\partial f_a}{\partial t}=\pi (2\pi)^{d}\frac{\partial}{\partial v_i}  \int d{\bf k} \, d{\bf v}'
\, k_ik_j  \frac{\hat{u}(k)^2}{|\epsilon({\bf k},{\bf k}\cdot {\bf v})|^2}\delta\lbrack {\bf k}\cdot
({\bf v}-{\bf v}')\rbrack \sum_b\left (m_b f'_b \frac{\partial f_a}{\partial {v}_{j}}-m_a f_a
\frac{\partial f'_b}{\partial {v'}_{j}}\right )
\label{ms1}
\end{equation}
with
\begin{eqnarray}
\epsilon({\bf k},\omega)=1+(2\pi)^{d}\hat{u}({k})\sum_b \int \frac{{\bf k}\cdot
\frac{\partial f_b}{\partial {\bf v}}}{\omega-{\bf k}\cdot {\bf v}}\, d{\bf v},
\label{ms1b}
\end{eqnarray}
where $f_a({\bf v},t)$ is the distribution function of species
$a$ normalized such that $\int f_a\, d{\bf v}=n_a m_a$ and the
sum $\sum_b$ runs over all the species. We can
use this equation to give a new interpretation of the test particle
approach developed in Sec. \ref{sec_relaxtest}. We make three assumptions:
(i) We assume that the system is composed of two types of particles, the
test particles with mass $m$ and the field particles with mass $m_f$; (ii) we
assume that the number of test particles is much lower than the number of
field particles; (iii) we assume that the field particles are in a steady
distribution $f({\bf v})$. Because of assumption (ii), the
collisions between the field particles and the test particles do not alter the
distribution of the field particles so that the field particles remain in
their steady state. The collisions of the test particles among themselves
are also negligible, so they only evolve due to the collisions
with the
field particles. Therefore, if we call $P({\bf v},t)$ the
distribution function of the test particles (to have notations similar to
those of Sec. \ref{sec_relaxtest} with, however, a different interpretation),
its evolution is given by the Fokker-Planck equation obtained from Eq. (\ref{ms1}):
\begin{equation}
\frac{\partial P}{\partial t}=\pi (2\pi)^{d}\frac{\partial}{\partial v_i}  \int d{\bf k} \, d{\bf v}'
\, k_ik_j  \frac{\hat{u}(k)^2}{|\epsilon({\bf k},{\bf k}\cdot {\bf v})|^2}\delta\lbrack {\bf k}\cdot
({\bf v}-{\bf v}')\rbrack \left (m_f f' \frac{\partial P}{\partial {v}_{j}}-m P
\frac{\partial f'}{\partial {v'}_{j}}\right )
\label{ms2}
\end{equation}
with
\begin{eqnarray}
\epsilon({\bf k},\omega)=1+(2\pi)^{d}\hat{u}({k})\int \frac{{\bf k}\cdot
\frac{\partial f}{\partial {\bf v}}}{\omega-{\bf k}\cdot {\bf v}}\, d{\bf v}.
\label{ms2b}
\end{eqnarray}
The diffusion tensor is given by
\begin{eqnarray}
D_{ij}=\pi(2\pi)^d m_f \int d{\bf k}\, d{\bf v}' \, k_i k_j \frac{\hat{u}(k)^2}{|\epsilon({\bf k},{\bf k}\cdot {\bf v})|^2}\delta\lbrack {\bf k}\cdot ({\bf v}-{\bf v}')\rbrack f({\bf v}'),
\label{ms3}
\end{eqnarray}
and the friction force due to the polarization by
\begin{eqnarray}
{F}_i^{pol}=\pi (2\pi)^d m\int d{\bf k}\, d{\bf v}'\, k_i k_j \frac{\hat{u}(k)^2}{|\epsilon({\bf k},{\bf k}\cdot {\bf v})|^2}\delta\lbrack {\bf k}\cdot ({\bf v}- {\bf v}')\rbrack \frac{\partial f}{\partial {v'_j}}({\bf v}').
\label{ms4}
\end{eqnarray}
We recall that the diffusion coefficient is due to the fluctuations of the  force produced by the field
particles, while the friction by polarization is due to the
 perturbation on the distribution of the field particles caused by the test particles. This explains
 the occurrence of the masses $m_f$ and $m$ in Eqs. (\ref{ms3}) and (\ref{ms4}) respectively.

If we neglect collective effects, using Eq. (\ref{test8}) and noting that \cite{landaud}:
\begin{eqnarray}
\label{ms5}
\frac{\partial D_{ij}}{\partial v_j}=\frac{m_f}{m}F^{pol}_i,
\end{eqnarray}
we get
\begin{equation}
\label{ms6}
{\bf F}_{friction}=\left (1+\frac{m_f}{m}\right ){\bf F}_{pol}.
\end{equation}
If we assume furthermore that $m\gg m_f$, we find that ${\bf
F}_{friction}\simeq {\bf F}_{pol}$. However, in general, the friction
force is different from the friction by polarization. The other
results of Sec. \ref{sec_relaxtest} can be easily generalized to
multi-species systems. The special features of the dimension
$d=1$
for multi-species systems is discussed in Sec. 7 of \cite{cvb}.

If the field particles have an isothermal distribution, then from
Eq. (\ref{ms4}),  ${F}_i^{pol}=-D_{ij}({\bf v})\beta m v_j=-D_{\|}({\bf v})\beta
m v_i$ and the Fokker-Planck equation (\ref{ms2}) reduces to
\begin{equation}
\label{tb5app}\frac{\partial P}{\partial t}=\frac{\partial}{\partial v_{i}} \left \lbrack D_{ij}({\bf v})\left (\frac{\partial P}{\partial
v_{j}}+\beta m  P v_j\right )\right\rbrack,
\end{equation}
where $D_{ij}({\bf v})$ is given by Eq. (\ref{ms3}) determined by the field particles. If the distribution of the field
particles is $f\propto e^{-\beta m_f v^2/2}$, the equilibrium distribution of the test particles is $P\propto
e^{-\beta m v^2/2}\propto f^{m/m_f}$. When $m\gg m_f$, the evolution is dominated by frictional effects;
when $m\ll m_f$ it is dominated by diffusion. The relaxation time scales like $t_{R}^{bath}\sim v_m^2/D$
where $v_m^2$ is the r.m.s. velocity of the test particles (with mass $m$) and $D$ is determined by the
field particles (with mass $m_f$). Therefore, $t_{R}^{bath}\sim 1/(D\beta m)\sim 1/\xi$ where $\xi$ is the
friction coefficient associated to the friction by polarization. If we neglect
collective effects, the ``true'' friction coefficient is
$\xi_*=(1+m_f/m)\xi$ and the ``true'' friction
is ${F}_i^{friction}=-D_{ij}({\bf v})\beta (m+m_f) v_j=-D_{\|}({\bf v})\beta
(m+m_f) v_i$.

\section{Dimensionless Fokker-Planck equation}
\label{sec_diless}

With the change of variables ${\bf w}=(\beta m/2)^{1/2}{\bf v}$, the Fokker-Planck equation (\ref{tb5}) may be
rewritten as
\begin{equation}
\label{diless1}\frac{\partial P}{\partial t}=\frac{1}{2v_m^2}\frac{\partial}{\partial w_{i}} \left \lbrack
D_{ij}({\bf w})\left (\frac{\partial P}{\partial
w_{j}}+2 P w_j\right )\right\rbrack
\end{equation}
with
\begin{equation}
\label{diless2}D_{ij}({\bf w})=\int d\hat{\bf k}\, \hat{k}_i \hat{k}_j e^{-(\hat{\bf k}\cdot {\bf w})^2}{\cal D}_d(\hat{\bf k}\cdot {\bf w}).
\end{equation}
When collective effects are neglected, using $D_{ij}({\bf w})={\cal D}_d G_{ij}({\bf w})$ it may be written as
\begin{equation}
\label{diless3}\frac{\partial P}{\partial t}=\frac{1}{t_R}\frac{\partial}{\partial w_{i}} \left \lbrack
G_{ij}({\bf w})\left (\frac{\partial P}{\partial
w_{j}}+2 P w_j\right )\right\rbrack
\end{equation}
with
\begin{equation}
\label{diless4}G_{ij}({\bf w})=\int d\hat{\bf k}\, \hat{k}_i \hat{k}_j e^{-(\hat{\bf k}\cdot {\bf w})^2},
\qquad t_R
=\frac{2v_m^2}{{\cal D}_d}.
\end{equation}
These scalings clearly show the emergence of the relaxation timescale $t_R$.


\begin{thebibliography}{}


\bibitem{houches}  {\small  {\it Dynamics and Thermodynamics of Systems with Long-Range Interactions}, edited by T. Dauxois, S. Ruffo, E. Arimondo and  M. Wilkens, Lectures  Notes in Physics {\bf 602} (Berlin: Springer, 2002)}
\bibitem{assise}  {\small {\it Dynamics and Thermodynamics of Systems with Long-Range
Interactions: Theory and Experiments}, edited by A. Campa, A. Giansanti, G. Morigi and F. Sylos Labini, AIP Conf. Proc. {\bf 965} 122 (2008)}
\bibitem{oxford}  {\small  {\it Long-Range Interacting Systems}, edited by T. Dauxois, S. Ruffo and L. Cugliandolo, Les Houches Summer School 2008, (Oxford: Oxford University Press, 2009)}
\bibitem{cdr}  {\small A. Campa, T. Dauxois, S. Ruffo,   Physics Reports {\bf 480}, 57 (2009)}
\bibitem{proceedingdenmark} {\small P.H. Chavanis, Theor. Comput. Fluid Dyn. {\bf 24}, 217 (2009)}
\bibitem{bgm} {\small F. Bouchet, S. Gupta, D. Mukamel, Physica A {\bf 389}, 4389 (2010)}
\bibitem{paper1}  {\small  P.H. Chavanis, Eur. Phys. J. Plus {\bf 127}, 19 (2012) [Paper I]}
\bibitem{paper3}  {\small  P.H. Chavanis, [arXiv:1303.1004] [Paper III]}
\bibitem{boltzmann}  {\small L. Boltzmann, Wien, Ber.  {\bf 66}, 275 (1872)}
\bibitem{chapman}  {\small S. Chapman, T.G. Cowling, {\it The Mathematical Theory of Non-Uniform Gases} (Cambridge University Press, 1952)}
\bibitem{landau}  {\small L.D. Landau, Phys. Z. Sowj. Union  {\bf 10}, 154 (1936)}
\bibitem{dh1}  {\small  P. Debye, E. H\"uckel, Physik Z. {\bf 24}, 185 (1923)}
\bibitem{dh2}  {\small  P. Debye, E. H\"uckel, Physik Z. {\bf 24}, 305 (1923)}
\bibitem{chandra}  {\small S. Chandrasekhar, {\it Principles of Stellar Dynamics} (University of Chicago Press, 1942)}
\bibitem{chandra1}  {\small S. Chandrasekhar, Astrophys. J.  {\bf 97}, 255 (1943) }
\bibitem{chandra2}  {\small S. Chandrasekhar, Astrophys. J.  {\bf 97}, 263 (1943) }
\bibitem{schwarzschild}  {\small  K. Schwarzschild, Seeliger Festschrift, 94 (1924)}
\bibitem{rosseland}  {\small S. Rosseland, Mon. Not. R. Astron. Soc.  {\bf 88}, 208 (1928)}
\bibitem{jeansbook}  {\small J. Jeans {\it Astronomy and Cosmogony} (Cambridge University Press, New York, 1929)}
\bibitem{smart}  {\small  W.M. Smart, {\it Stellar Dynamics} (Cambridge Univ. Press, 1938)}
\bibitem{spitzerevap}  {\small L. Spitzer, Mon. not. R. astron. Soc.  {\bf 100}, 396 (1940)}
\bibitem{ambart}  {\small V.A. Ambartsumian,  Ann. Leningrad State Univ. {\bf 22}, 19 (1938)}
\bibitem{cvn}  {\small S. Chandrasekhar, J. von Neumann,  Astrophys. J. {\bf 95}, 489 (1942)}
\bibitem{cohen}  {\small R.S. Cohen, L. Spitzer and P.M. Routly, Phys. Rev. {\bf 80}, 230 (1950) }
\bibitem{nice}  {\small S. Chandrasekhar, Rev. Mod. Phys. {\bf 21}, 383 (1949) }
\bibitem{chandrastoch}  {\small S. Chandrasekhar, Rev. Mod. Phys. {\bf 15}, 1 (1943) }
\bibitem{rosen}  {\small M. Rosenbluth, W. MacDonald, D. Judd,  Phys. Rev. {\bf 107}, 1 (1957)}
\bibitem{kingL}  {\small I. King, Astron. J. {\bf 65}, 122 (1960)}
\bibitem{kandrup1}  {\small H. Kandrup,  Astrophys. J. {\bf 244}, 316 (1981)}
\bibitem{aanew}  {\small  P.H. Chavanis, Astron. Astrophys. {\bf 556}, A93 (2013)}
\bibitem{gruner}  {\small  P. Gruner, Ann. der. Physik {\bf  35}, 381 (1911)}
\bibitem{chapmanplasma}  {\small  S. Chapman, Mon. Not. R. Astron. Soc. {\bf 82}, 5 (1922)}
\bibitem{persico}  {\small  E. Persico, Mon. Not. R. Astron. Soc. {\bf 86}, 93 (1926)}
\bibitem{ross}  {\small  S. Rosseland, Mon. Not. R. Astron. Soc. {\bf 84}, 720 (1924)}
\bibitem{edd}  {\small  A.S. Eddington, Mon. Not. R. Astron. Soc. {\bf 86}, 1 (1925)}
\bibitem{vlasov1}  {\small  A.A. Vlasov, Zh. Eksp. Teor. Fiz. {\bf  8}, 291 (1938)}
\bibitem{vlasov2}  {\small  A.A. Vlasov, J. Phys. (U.S.S.R.) {\bf  9}, 25 (1945)}
\bibitem{jeans}  {\small  J. Jeans, Mon. Not. R. Astron. Soc. {\bf 76}, 71 (1915)}
\bibitem{henon}  {\small  M. H\'enon, Astron. Astrophys.  {\bf 114}, 211 (1982)}
\bibitem{landaudamping}  {\small  L.D. Landau, J. Phys. (U.S.S.R.) {\bf  10}, 25 (1946)  }
\bibitem{lb}  {\small  D. Lynden-Bell, Mon. Not. R. Astron. Soc.  {\bf 136}, 101 (1967)}
\bibitem{levin}  {\small  Y. Levin, R. Pakter, F.B. Rizzato, Phys. Rev. E {\bf 78}, 021130 (2008)}
\bibitem{stan}  {\small  F. Staniscia, P.H. Chavanis, G. De Ninno, Phys. Rev. E {\bf 83}, 051111 (2011)}
\bibitem{pb}  {\small  D. Pines, D. Bohm, Phys. Rev.  {\bf 85}, 338 (1952)}
\bibitem{bogoliubov}  {\small  N.N. Bogoliubov, {\it Problems of Dynamical Theory in Statistical Physics},'' Section 11. Moscow, (1946)}
\bibitem{lenard}  {\small A. Lenard, Ann. Phys. (N.Y.) {\bf 10}, 390 (1960)}
\bibitem{balescu1}  {\small   R. Balescu, Rev. Mod. Phys. {\bf 32}, 719 (1960a)}
\bibitem{balescu2}  {\small   R. Balescu, Phys. Fluids {\bf 3}, 52 (1960b)}
\bibitem{priba}  {\small  I. Prigogine, R. Balescu, Physica {\bf  23}, 555 (1957)}
\bibitem{rostoker}  {\small  N. Rostoker, Phys. Fluids {\bf 7}, 479 (1964)}
\bibitem{liboff}  {\small  R.L. Liboff,  Phys. Fluids {\bf 2}, 40 (1959)}
\bibitem{guernseyphd}  {\small  R.L. Guernsey, PhD thesis, University of Michigan (1960) (unpublished)}
\bibitem{wu}  {\small T.Y. Wu, {\it Kinetic equations of gases and plasmas} (Addison-Wesley, 1966)}
\bibitem{baldwin}  {\small  D.E. Baldwin,  Phys. Fluids {\bf 5}, 1523 (1962)}
\bibitem{fb}  {\small  E.A. Frieman, D.L. Book,  Phys. Fluids {\bf 6}, 1700 (1963)}
\bibitem{weinstock}  {\small  J. Weinstock,  Phys. Rev.  {\bf 133}, A673 (1964)}
\bibitem{guernsey}  {\small  R.L. Guernsey,  Phys. Fluids {\bf 7}, 1600 (1964)}
\bibitem{ka}  {\small  T. Kihara, O. Aono,  J. Phys. Soc. Japan {\bf 18}, 837 (1963)}
\bibitem{gould}  {\small  H.A. Gould, H.E. DeWitt,  Phys. Rev.  {\bf 155}, 68 (1967)}
\bibitem{temko}  {\small  S.V. Temko,  Sov. Phys. JETP {\bf 4}, 898 (1957)}
\bibitem{kadomtsev}  {\small B.B. Kadomtsev,  Sov. Phys. JETP {\bf 6}, 117 (1958)}
\bibitem{tchen}  {\small  C.M. Tchen, Phys. Rev.  {\bf 114}, 394 (1959)}
\bibitem{ichikawa}  {\small  Y.H. Ichikawa, Progr. Theoret. Phys.  {\bf 24}, 1083 (1960)}
\bibitem{willis}  {\small  C.R. Willis,  Phys. Fluids {\bf 5}, 219 (1962)}
\bibitem{gasiorowicz}  {\small  S. Gasiorowicz, M. Neuman, R.J. Riddell,  Phys. Rev. {\bf 101}, 922 (1956)}
\bibitem{rr}  {\small  N. Rostoker, M.N. Rosenbluth,  Phys. Fluids {\bf 3}, 1 (1960)}
\bibitem{th}  {\small W.B. Thompson, J. Hubbard, Rev. Mod. Phys. {\bf 32}, 714 (1960)}
\bibitem{hubbard1}  {\small J. Hubbard, Proc. R. Soc. Lond. {\bf 260}, 114 (1961a)}
\bibitem{hubbard2}  {\small J. Hubbard, Proc. R. Soc. Lond. {\bf 261}, 371 (1961b)}
\bibitem{dupree1}  {\small  T.H. Dupree,  Phys. Fluids {\bf 4}, 696 (1961)}
\bibitem{ichimaru}  {\small S. Ichimaru, {\it Basic Principles of Plasma Physics} (Benjamin, Reading, MA, 1973)}
\bibitem{nicholson}  {\small D.R. Nicholson, {\it Introduction to Plasma Theory} (Krieger Publishing Company, Malabar, Florida, 1992)}
\bibitem{dupree2}  {\small  T.H. Dupree,  Phys. Fluids {\bf 6}, 1714 (1963)}
\bibitem{fried}  {\small B.D. Fried, in {\it Plasma Physics in Theory and Application}, edited by W.B. Kunkel (McGraw-Hill, New York, 1966)}
\bibitem{klimontovich}  {\small Y.L. Klimontovich, {\it The Statistical Theory of Non-Equilibrium Processes in a Plasma} (M.I.T. press, Cambridge, 1967)}
\bibitem{pitaevskii}  {\small  E.M. Lifshitz, L.P.  Pitaevskii, {\it Physical Kinetics} (Pergamon Press, Oxford, 1981)}
\bibitem{kac}  {\small M. Kac, G.E. Uhlenbeck, P.C. Hemmer, J. Math. Phys.  {\bf 4}, 216 (1963)}
\bibitem{angleaction}  {\small  P.H. Chavanis, Physica A {\bf 377}, 469 (2007)}
\bibitem{kindetail}  {\small  P.H. Chavanis, J. Stat. Mech (2010) P05019}
\bibitem{heyvaerts}  {\small  J. Heyvaerts, Mon. Not. R. Astron. Soc. {\bf 407}, 355 (2010)}
\bibitem{newangleaction}  {\small  P.H. Chavanis, Physica A {\bf 391}, 3680 (2012)}
\bibitem{kinonsager}  {\small  P.H. Chavanis, Physica A {\bf 391}, 3657 (2012)}
\bibitem{feix}  {\small O.C. Eldridge, M. Feix,  Phys. Fluids.  {\bf 6}, 398 (1962) }
\bibitem{kp}  {\small B.B. Kadomtsev, O.P. Pogutse, Phys. Rev. Lett.  {\bf 25}, 1155 (1970) }
\bibitem{bd}  {\small F. Bouchet, T.  Dauxois,  Phys. Rev. E {\bf 72}, 045103 (2005)}
\bibitem{cvb}  {\small  P.H. Chavanis, J. Vatteville, F.  Bouchet,  Eur. Phys. J. B {\bf  46}, 61 (2005)}
\bibitem{dawson}  {\small J. Dawson,  Phys. Fluids  {\bf 7}, 419 (1964)}
\bibitem{rouetfeix}  {\small J.L. Rouet, M. Feix,  Phys. Fluids B  {\bf 3}, 1830 (1991)}
\bibitem{campa}  {\small A. Campa, A.  Giansanti, G.  Morelli, Phys. Rev. E {\bf 76}, 041117 (2007)}
\bibitem{barre1}  {\small J. Barr\'e, A. Olivetti, Y.Y. Yamaguchi, J. Stat.
Mech. (2010) P08002}
\bibitem{barre2}  {\small J. Barr\'e, Y.Y. Yamaguchi, J. Phys. A {\bf 44},
405502  (2011)}
\bibitem{barre3}  {\small J. Barr\'e, Y.Y. Yamaguchi, J. Phys. A {\bf 46},
225501  (2013)}
\bibitem{joyce} {\small M. Joyce, T. Worrakitpoonpon, J. Stat. Mech. {\bf 10}, 12 (2010)}
\bibitem{yamaguchi}  {\small Y. Yamaguchi, J.   Barr\'e, F. Bouchet, T.  Dauxois, S. Ruffo,   Physica A {\bf 337}, 36 (2004)}
\bibitem{campaall}  {\small A. Campa, P.H. Chavanis, A.  Giansanti, G. Morelli,  Phys. Rev. E {\bf 78}, 040102 (2008) }
\bibitem{risken}  {\small H. Risken, {\it The Fokker-Planck equation} (Springer, 1989)}
\bibitem{kalnajs}  {\small  A. Kalnajs, Astrophys. J. {\bf 166}, 275 (1971)}
\bibitem{kandrup2}  {\small H. Kandrup,   Astro. Space. Sci.  {\bf 97}, 435 (1983)}
\bibitem{hb4}  {\small  P.H. Chavanis, Physica A {\bf 387}, 1504 (2008)}
\bibitem{landaud}  {\small  P.H. Chavanis,  Eur. Phys. J. B {\bf  52}, 61 (2006)}
\bibitem{pre}  {\small P.H. Chavanis, Phys. Rev. E {\bf 64}, 026309 (2001)}
\bibitem{hb2}  {\small P.H. Chavanis, Physica A  {\bf 361}, 81 (2006)}
\bibitem{hb1}  {\small P.H. Chavanis, Physica A  {\bf 361}, 55 (2006)}
\bibitem{bt}  {\small  J. Binney, S.  Tremaine, {\it Galactic Dynamics} (Princeton Series in Astrophysics, 1987)}
\bibitem{heggie}  {\small  D.C. Heggie, Mon. Not. R. Astron. Soc.  {\bf 173}, 729 (1975)}
\bibitem{nr}  {\small T. O'Neil, N. Rostoker,  Phys. Fluids  {\bf 8}, 1109 (1965)}
\bibitem{ti}  {\small H. Totsuji, S. Ichimaru, Prog. Theor. Phys.  {\bf 50}, 753 (1973)}
\bibitem{reponselin}  {\small P.H. Chavanis, Eur. Phys. J. Plus {\bf 128}, 38
(2013)}
\bibitem{stahl}  {\small B. Stahl, M.K.H. Kiessling, K. Schindler, Planet. Space Sci. {\bf 43}, 271
(1994)}
\bibitem{exclusion}  {\small P.H. Chavanis [arXiv:1309.2843]}
\bibitem{nardini1}  {\small C. Nardini, S. Gupta, S. Ruffo, T. Dauxois, F. Bouchet, J. Stat. Mech. (2012) L01002}
\bibitem{nardini2}  {\small C. Nardini, S. Gupta, S. Ruffo, T. Dauxois, F. Bouchet, J. Stat. Mech.
{\bf 12}, 12010 (2012)}
\bibitem{gardiner}  {\small C.W. Gardiner, {\it Handbook of Stochastic Methods} (Springer, 1990)}
\bibitem{llif}  {\small L. Landau, E. Lifshitz {\it Fluid Mechanics} (Pergamon, London, 1959)}
\bibitem{hb5}  {\small P.H. Chavanis, Physica A  {\bf 387}, 5716 (2008)}
\bibitem{dean}  {\small D.S. Dean, J. Phys. A  {\bf 29}, L613 (1996)}
\bibitem{lc}  {\small M. Lemou, P.H. Chavanis, Physica A   {\bf 389}, 1021 (2010)}



















\end{thebibliography}
\end{document}